> "Who controls the past controls the future.
> Who controls the present controls the past."
> – George Orwell, 1984

# Chapter 9   History of gradient advances in SRF[1]

*H. Padamsee, Cornell University and Fermilab*

## 9.1   Introduction

In this chapter, I will attempt to the best of my knowledge to trace the history of gradients evolution in the field of superconducting radio frequency (SRF). I will restrict the scope to primary innovations along with some of the ensuing developments. But I will not cover all the many applications and findings over the subsequent decades of progress that were based on the primary discoveries and inventions. I will also not cover a number of other important topics in the history of cavity developments, such as the drive for higher $Q$ values, or the push for lower cavity costs (via Nb/Cu cavities or large grain Nb). Other interesting topics left out are $Nb_3Sn$ and high $T_c$ superconductors.

Many aspects of SRF development are thoroughly covered in two texts [1-2], many review papers [3-4], and most completely in the proceedings of International SRF Conferences [5].

Fig. 9.1 shows the evolution of $Q_0$ vs $E$ curves over the history of the field along with performance limitations from the phenomena to be discussed throughout the chapter: multipacting, thermal breakdown (quench), field emission, high field $Q$-slope and the hydrogen $Q$-disease. The figure also shows the ideal $Q_0$ vs $E$ curve that we expect if there are no limitations (this is rarely the case in practice), and the ideal performance if the theoretical capability of Nb were to be reached to obtain the ultimate accelerating field.

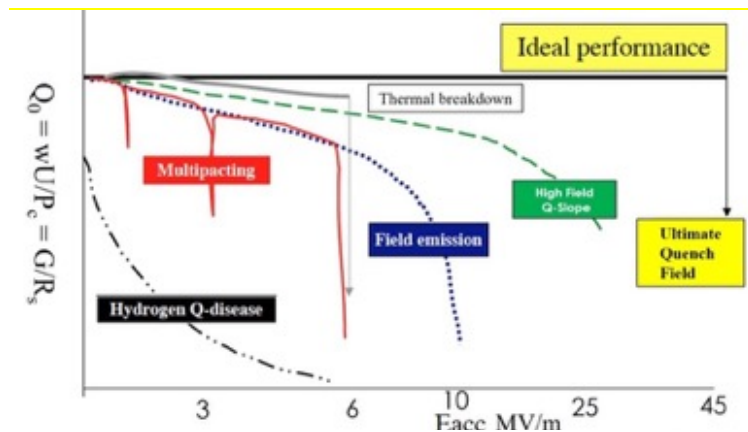

**Figure 9.1:** $Q_0$ vs $E$ curves showing the ideal performance of Nb and the typical $Q$-degradations due to various phenomena. The ideal performance is a flat $Q_0$ curve till the ultimate field is reached at the theoretical superheating field limit $H_{sh}$.

---

[1] This chapter will be part of the forthcoming book *"Radio frequency superconductivity for modern accelerators: Reference book for scientists and engineers,"* ed. by S. Belomestnykh, A. Grassellino, ansd A. Romanenko

## *9.2 Summary of historical gradient advances*

Over the last six decades there have been several breakthroughs in gradients either following or accompanying a leap forward in the fundamental understanding of the limiting mechanisms that prevented the rise of gradients. (References to the major contributors for these developments are given in the appropriate sections.) Before 1980, the dominant phenomenon limiting cavity gradients to about 3 MV/m (e.g., with 1.3 GHz multicell cavities) was one-surface multipacting (MP). Multipacting is a resonant process in which an electron avalanche builds up within a small region of the cavity surface due to a confluence of several circumstances, as explained later. This phenomenon is described in detail in Chapter 8. The invention of the round wall (spherical) cavity shape provided the best solution to the one-surface MP problem, finally opening the door to expanding the gradient frontier. Several other cures to MP were also explored, which we will discuss, but these did not enjoy the success of the revolutionary shape choice. The initial spherical choice was improved for mechanical stability by the elliptical cell profile which also provided more effective drainage of etching and rinsing liquids. Elliptical cavities are now commonly adopted for $\beta = 1$ applications, such as European XFEL [6-8] and LCLS-II [9].

Despite the solution of the MP problem by the spherical/elliptical shape, gradients only rose to about 4 – 6 MV/m due to breakdown of superconductivity at defects, with a so-called "quench". When sub-millimeter-size regions with high RF losses (called defects) heat up in the RF field, the temperature of the good superconductor just outside the defect rises. With increasing gradient, the temperature of the superconductor near the defect eventually exceeds the superconducting transition temperature $T_c$. RF losses around the defect increase substantially, leading to thermal runaway, and a "quench" of superconductivity over a large (> 1 cm$^2$) region. The best solution to mitigate thermal breakdown was to use higher thermal conductivity Nb via higher purity Nb, characterized by the residual resistivity ratio (RRR). Again, there were other cures which will also be discussed. With the higher thermal conductivity, a given defect can tolerate more RF dissipation at higher fields before driving the neighboring superconductor into the normal state. The best solution for high purity, high RRR Nb was to encourage Nb producing industry to improve their electron beam melting furnaces and practices. With higher RRR Nb, average gradients rose to about 10 MV/m when field emission (FE) of electrons took over as the dominant limitation. The phenomenon of field emission is described in Chapter 8.

Briefly, during field emission, the $Q_0$ of a niobium cavity starts to fall exponentially with increasing electron currents emerging from particular emitting spots on the surface. Research on the origin of field emission showed that micro-particle contaminants are the dominant sources. High pressure water rinsing (HPR) and cavity assembly in class 10 – 100 clean room environments, along with high levels of cleanliness in cavity surface preparation, led to fewer emission sites and accompanying improvements in cavity gradients to about 20 MV/m.

Above 20 – 25 MV/m, a new phenomenon, called the High Field Q-slope (HFQS) took over, dropping the cavity $Q_0$ exponentially with field. The complete understanding of HFQS is still in progress with a reasonable model in hand, to be discussed later in this chapter. HFQS is also addressed in Chapter 4 in more detail. But the empirical cure to HFQS was discovered quickly. Two steps are essential – prepare smooth surfaces by EP and bake the cavity at 120ºC for about two days. Without the crucial baking step, the HFQS continued to dominate gradient limits for cavities prepared by most methods. Soon gradients rose to 30 – 35 MV/m with record values 40 – 45 MV/m in 9-cell 1.3 GHz cavities.

To circumvent the hard barrier of the fundamental critical field, the next advance (after curing the HFQS with EP and 120ºC baking) came from altering the cavity shape so that the surface magnetic field in the cavity structure would be lower by 10 – 15% for the same accelerating field. This was achieved by increasing the surface area of the cavity near the equator to lower the current density and the peak magnetic field there. To get the most reduction of the surface magnetic field the shapes chosen allowed the peak electric field to increase by 20% with the thought that field emission due to higher $E_{pk}$ could be dealt with by better rinsing methods, whereas the fundamental critical magnetic field presented a hard limit. Starting with the Re-entrant shape, the Low-Loss shape and the Ichiro shape were invented, and single cell cavities reached gradients of 50 – 59 MV/m at high $Q$. However multi-cell cavities of these shapes have not yet been able to achieve single cell performance levels, mostly due to the higher field emission from the higher $E_{pk}$. A new shape, called Low Surface Field (LSF) is now under exploration where $H_{pk}$ is 15% lower but the $E_{pk}/E_{acc}$ is not increased above 2.0, its canonical shape value.

Very recently (after 2010), another avenue for higher gradients came via the nitrogen-doping (N-doping) method, which was invented mainly to raise cavity $Q$ values, especially at medium gradients (15 – 20 MV/m) for CW operation. For high gradients, a variation of N-doping, called N-infusion, came into play to ameliorate HFQS. After 800ºC heat treatment (to remove hydrogen absorbed by niobium during chemical treatment), the temperature in the furnace was reduced to 120ºC and 25 mtorr of $N_2$ introduced. The HFQS was not only removed, but higher fields and higher $Q$'s became regularly possible up to 40 – 45 MV/m.

Yet another very recent cure was "two-step baking" which proved even more effective against the HFQS to allow accelerating fields near 50 MV/m. These recent developments are covered more thoroughly in Chapter 4. At this level, the surface magnetic fields approach the fundamental superheating critical magnetic field of 220 mT, and quenches could be triggered by very small imperfections. Even the possibility of direct magnetic phase transitions to the normal state has become an important candidate. At this stage, the superheating critical field of Nb presents a hard barrier to further gradient advances.

An overarching model has been proposed to explain different treatments to overcome HFQS. A thin hydrogen-rich layer (estimated 1 – 20 atomic %) exists near surface. Nb-H precipitates form in this hydrogen-rich layer at favorable nucleation sites. At $H_{pk}$ ~ 100 mT the largest hydride precipitate starts to transition to the normal conducting state, to manifest as the onset of the HFQS. As the field rises the smaller hydrides turn normal. The 120ºC baking and N infusion cures to HFQS are explained by the injection and diffusion of O or N from the surface into the RF layer. The interstitials injected serve as trapping centers to prevent H from diffusing freely to form hydrides.

Key to the many gradient advances was the accompanying development of thermometry-based diagnostic techniques to detect the source of the limitations. We will also discuss the advances of these important diagnostic methods. To be complete, we briefly discuss better manufacturing methods and better surface preparation methods that played essential roles in the steady march of niobium cavity gradients. Accordingly we will discuss sheet metal hydroforming, stamping (deep-drawing) or spinning methods to replace machining cavities from solid niobium, better electron beam welding methods to avoid weld defects, better surface processing methods, such as buffered chemical polishing and electropolishing to remove surface damage and provide smooth surfaces, better final treatment procedures, such as high pressure water rinsing, to remove chemical other contaminants, and clean assembly techniques in Class 10 – 100 clean rooms to avoid FE causing dust contamination. Annealing cavities in furnaces at 800 – 1000ºC played a significant role to remove hydrogen which damaged cavity $Q$ values, as well as final baking at 120ºC all worked together to overcome many types of gradient limitations (and $Q$). Some of the more recent developments are still maturing to reach improved understanding and application to multi-cell cavities.

## 9.3  Pioneering developments in SRF

Fairbanks, Wilson, and Schwettman [10-11] at High Energy Physics Laboratory (HEPL) at Stanford University pioneered the development of SRF from the early 1960s through the early 1970s with the goal of building an SRF electron linac at 2 GeV to achieve the scientific goals of nuclear physics [12-13]. The target was a gradient of 14 MV/m with CW operation and with $Q$ values of several times $10^9$, to keep refrigeration loads below 10 kW (at 2 K), which would be a major cryogenics challenge, and is even so at present times. With amazing foresight, Todd Smith at HEPL [14] introduced a backup plan to reach 2 GeV with a 3-pass recirculating linac which could operate at a modest gradient of 5 – 6 MV/m and $Q$ values near $10^9$. To realize recirculation, it was necessary to address beam breakup by damping cavity higher order modes, but this topic is beyond the scope of this chapter.

The landmark SRF achievement for Nb cavities came at HEPL [13, 15] from Turneaure and Viet around 1970 with $TM_{010}$ Nb cavities at 8.6 GHz (X-band). The cavities (Fig. 9.2) were machined in two halves from solid Nb, then joined with an electron beam (EB) weld at the center.

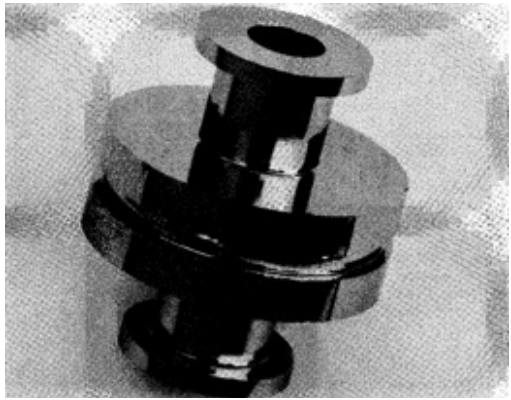

**Figure 9.2:** HEPL X-band 8.6 GHz cavity which reached high gradients at high $Q$'s [15].

It was a first and successful introduction of electron beam welding as a fabrication technique for SRF. With the goal of high performance, they fired the cavities in a UHV furnace at 1750 – 2100ºC. followed by chemical etching and a second UHV firing. The cavities showed spectacular performance (for the time) with a peak RF magnetic field of 108 mT, and $E_{pk}$ of 70 MV/m. At these peak fields, a $Q_0$ of $8 \times 10^9$ was measured. These values are close to those regularly achieved in multicell 1.3 GHz cavities today, a great tribute to the pioneering result. As another first, they also detected Lorenz-force detuning at these high fields. After this ground-breaking success, Nb became the material of choice for electron linacs, as opposed to lead-plated resonators, which HEPL and other labs were pursuing.

Encouraged by the high accelerating field in the 8.6 GHz cavity, HEPL quickly embarked upon 1.3 GHz Nb cavities more suitable for their CW electron linac ambitions. The jump to 1.3 GHz based on 8.6 GHz successful results was unfortunate, as would become clear later when it was discovered that multipacting would become the dominant limiting phenomenon, and that troublesome multipacting field levels scale with RF frequency. By the end of 1973, several full-length 55-cell superconducting structures at 1.3 GHz were built, tested and installed in the linac [16-17]. These were assembled from 7-cell substructures, made from hydroformed half-cells, as shown in Fig. 9.3. HEPL had moved in the 1970's from machining cavities from solid niobium to the less expensive method of hydroforming half-cells, subsequently joined by electron beam welding.

The initial operation of these structures [16] produced energy gradients from 2.0 to 3.8 MV/m and $Q$ values from 2 to $6\times10^9$. This was far short of the 14 MV/m goal and even the 5 MV/m goal for a recirculating linac. The inability to make progress in achieving high gradients was a crushing blow for electro-nuclear physicists wanting a CW electron linac in the 2 GeV range. By 1977, HEPL installed parts of the Superconducting Linear Accelerator (SCA) consisting of a superconducting injector and four superconducting accelerator sections each 5.65 meters long and one orbit of recirculation to reach a beam energy of 84 MeV [18]. The SRF structures operated at gradients between 1.5 – 2.5 MV/m.

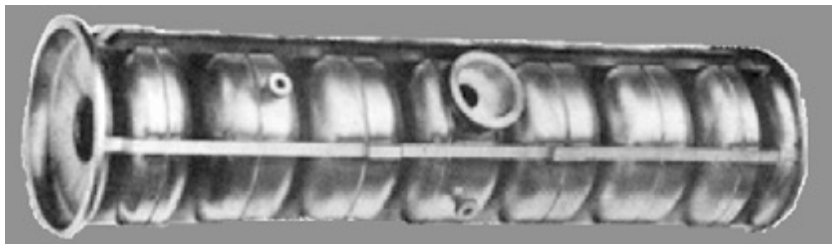

**Figure 9.3:** 1.3 GHz 7-cell substructure for the HEPL electron linac.

At Cornell, SRF work began in 1969 when Tigner developed a new cavity design for the Cornell high-energy (12 GeV) electron synchrotron to achieve energies in the neighborhood of 25 GeV in an electron synchrotron of the size of Cornell machine, RF peak powers of the order of tens of megawatts would be required if copper accelerating cavities were to be used. SRF cavities with gradients above 3 MV/m offered the possibility of reducing the power by several orders of magnitude. Tigner's idea was to use a cavity in the shape of a rectangular muffin tin, with an upper and lower half, separated by an open midplane so that synchrotron radiation from the ring would escape through the empty midplane between the two halves, without hitting the superconducting surface and causing damage.

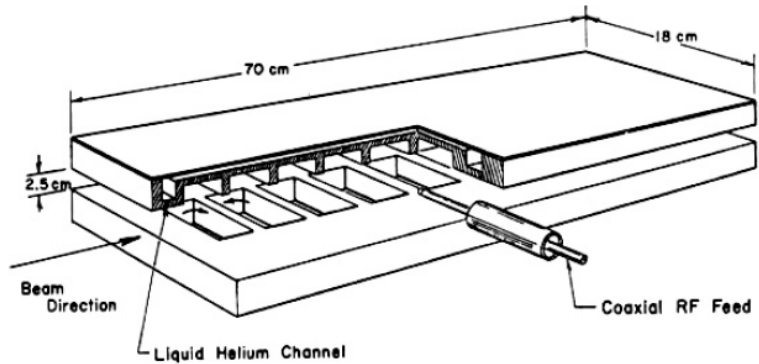

**Figure 9.4:** Cornell S-band 11-cell muffin-tin cavity [20].

Single cell and 11-cell muffin-tin cavities at 2.86 GHz (S-band) were machined by Kirchgessner from reactor grade solid Nb. Single cell muffin-tin cavities, when tested by Sundelin and Padamsee, also suffered multipacting [19] but, because of the higher frequency, achieved gradients of 4 – 6 MV/m with $Q$-values of the order of $10^9$.

In 1973, Sundelin and Tigner led development of a 60-cm-long niobium cavity. The 11-cell cavity (Fig. 9.4) was machined out of solid Nb and installed in the Cornell Electron Synchrotron. The cavity was tested by Sundelin et al. to accelerate a 4 GeV beam in 1974 [20]. Its performance was limited to 4 MV/m by multipacting and thermal breakdown. The $Q_0$ maintained its initial value of $1.1 \times 10^9$. This was the first application of SRF to a high energy physics accelerator.

Motivated by obvious practical considerations of economy and ease of fabrication, Kirchgessner [21] developed fabrication of muffin-tin half-cells at Cornell by deep drawing sheet metal in 1977 (the method is reviewed in Chapter 14) instead of hydroforming developed earlier at HEPL for SCA structures.

With the discovery of the Charm Quark in 1975 at electron-positron storage ring SPEAR at SLAC [22], many high energy physics laboratories, Cornell, KEK, and CERN became interested in building or upgrading $e+e-$ storage rings for particle physics. Cornell built CESR (5 GeV) in 1979, KEK's ambition was TRISTAN (25 GeV) commissioned in 1987, and CERN proposed LEP (45 GeV) commissioned in 1989. DESY's ambition was an electron-proton collider HERA with an electron storage ring of energy 28 GeV. For cost-effective higher energy electron synchrotrons and electron-positron storage rings, CW SRF operation at high gradients and high $Q$'s would be very important.

In 1973, Kojima from Tohoku University visited HEPL to learn the technology for SRF cavities and accelerators. Kojima had been responsible for electron linac development. After return to Japan from HEPL, Kojima established a small SRF group at KEK. The group studied fabrication and treatment procedures on C-band (6 GHz) single and multi-cell cavities. In single cell cavities $Q$ values of $2 \times 10^{10}$ and $E_{acc} = 10$ MV/m were achieved in 1979 [23]. In an acceleration test with a 9-cell cavity they demonstrated $E_{acc} = 3$ MV/m. As we will see, the selection of the high frequency was a wise choice.

By the end of 1979, KEK focused efforts on the possibility of increasing the energy of TRISTAN from 28.5 GeV to above 30 GeV [24] by adding SRF cavities to the existing normal conducting

500 MHz system of 104 nine-cell cavities. By 1980, Lengeler and Bernard at CERN established an SRF program to double the energy of LEP in a future upgrade [25]. LEP was formally approved in 1981 as an electron positron colliding beam ring with 90 GeV in the center of mass.

## 9.4  Understanding multipacting

Ben-Zvi and Turneaure in 1973 [26] launched experimental studies on electron loading in 1.2 – 1.3 GHz cavities that focused on field emission with X-ray measurements and X-ray photographs. Also, computer simulations including electron multiplication by secondaries presented strong hints about the location of multipactor trajectories shown in Fig. 9.5(a). Large quantities of electrons tend to form a *low-energy electron cloud near the major diameter of the cavity*. Fig. 9.5(a) shows an electron avalanche which is developing a low-energy electron cloud, hinting at the location of multipacting. Fig. 9.5(b) shows an X-ray photograph from the bombardment of the iris regions with field emitted electrons when the cavity operated close to $E_{pk}$ = 10 MV/m.

They also realized a simple and important scaling law for electron dynamics – electrons follow the same trajectory with the same energy at each point, if the electric field divided by the cavity frequency is the same [14].

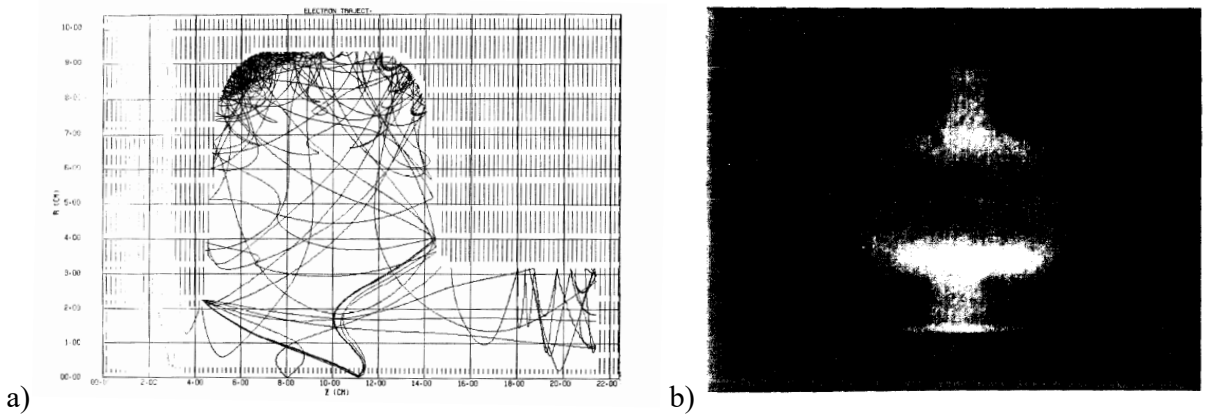

a)                                                                                              b)

**Figure 9.5:** (a) A typical computer plot of electron trajectories in a 1.3 GHz cavity of ten initial electrons and their second and higher generation electrons produced by the electron multiplication simulation program. (b) A photograph of X-rays from field emitted electrons bombarding the iris region of a 1.2 GHz cavity operating at $E_{pk}$ = 10 MV/m [26].

In the 1.3 GHz, 55-cell structures for the linac, field limitations came from quench at much lower gradients, for which field emission was less likely. This realization prompted development of diagnostics for thermal mapping to identify which of the seven substructures suffered quench. Researchers developed a novel technique based on second-sound propagation in superfluid helium [14]. They used an array of 14 resistance thermometers distributed along the length of the 7-cell structure to measure the time of arrival of the heat pulse initiated by breakdown. The point of origin could be established within ± 1 cell of the structure. With single-cell cavities and 7-cell structures, they mounted resistance thermometers on an arm which could be rotated azimuthally around the cavity axis to determine the location of the hot spot. They found breakdown to occur on the bottom surface

(equator). A thorough discussion of the evolution of temperature mapping diagnostics is found in Section 9.7.

In 1975, Schwettman [27] observed a crucial aspect of electron loading from seven-cell cavity results. Power absorption levels obeyed simple interval rule which could be explained by different order multipacting levels for some basic type of multipacting trajectory. This was the strong clue that electron multipacting was involved in limiting gradients at 1.3 GHz. Further, most of the 7-cell structures broke down at the same field level, suggesting that a single strong multipacting level was involved in initiating thermal breakdown.

To understand the low gradients (1.5 – 2.5 MV/m) at 1.3 GHz, HEPL scientists launched studies of superconducting cavities at the intermediate frequency of about 3 GHz (S-band), which allowed them to reach higher fields than at 1.3 GHz [28]. Peak electric fields of 35 MV/m in single cells were reached as compared to maximum of 22 MV/m in 1.3 GHz cavities. Following the hints about multipacting from seven-cell tests, Lyneis in 1977 used thermal mapping on a single-cell S-band cavity (Fig. 9.6(b)) operating at 1.4 K to locate the end point of the presumed multipacting trajectories. Using 100-Ohm carbon resistors attached to the cavity wall, he observed heat produced by the impacting multipacting electrons at the radius of the curvature between the outer cylindrical wall and the end wall of the cavity. Each resistor indicated in Fig. 9.6(a) represents a ring of resistors which are connected in series and span all azimuthal angles. Multipacting was observed at a number of discrete field levels. Each of these discrete MP levels was observable over a small field region, and within its field region, the MP intensity increased monotonically with incident rf power until the multipacting abruptly ceased, due to processing. The axial electric field $E_a$ at which the multipacting intensity was observed to reach a maximum is given in Table 9.1.

**Table 9.1:** Electron multipactor field levels for $TM_{010}$-mode S-band cavity [28].

| $E_a$ (exp) [MV/m] | $E_a$ (simul) [MV/m] | Order | $\overline{U}_{impact}$ [eV] | $\overline{\delta U}_{impact}$ | $F_{se}$ |
|---|---|---|---|---|---|
| 6.0 | 6.01 | 6 | 50 | 0.38 | 2.8 |
| 7.7 | 7.25 | 5 | 58 | 0.43 | 2.4 |
| 8.8 | 9.25 | 4 | 73 | 0.52 | 1.8 |
| 11.8 | 12.4 | 3 | 122 | 0.78 | 1.4 |
| 17.1 | 18.6 | 2 | 265 | 1.15 | 1.0 |

Once the impact location of the MP trajectories was established, Turneaure's electron trajectory calculation program identified the electron trajectories as one-point multipacting, driven by the small perpendicular electric field component at the radius of the curvature. Fig. 9.6(c) shows the calculated trajectories for 3$^{rd}$ order MP in agreement with the field level at which the cavity showed MP induced quenches. The name "one-point MP" described how the trajectory of the electron returns to near the point of origin. Because the return point is not exactly the same as the origin, the name could also be "one-surface MP". Electrons following the MP trajectories gain energy primarily from the perpendicular electric field component at the surface. Note how the selected multipacting trajectories of Fig. 9.6(c) concur with the earlier electron multiplication simulation of Fig. 9.5(a). Turneaure's simulation program (Fig. 9.5 (a)) was prophetic about the location of multipacting electrons in 1.3 GHz cavities.

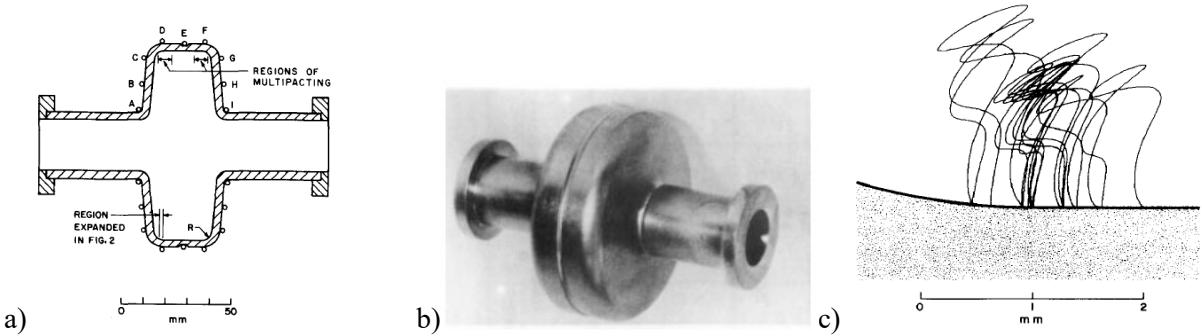

**Figure 9.6:** (a) Schematic drawing of the HEPL cavity showing locations of carbon resistors [28]. (b) Photograph of the cavity. (c) Trajectories of 3$^{rd}$ order MP [28].

Following the clues from Lyneis, Padamsee in 1977 [29] confirmed with a 100-Ohm carbon-resistor-based thermometry array that multipacting induced heating was taking place near the center of the muffin-tin cavity bottom, equivalent to the equator of the cylindrically symmetric HEPL pillbox geometry. These maps will be discussed in Section 9.7. The typical temperature fluctuations during multipacting measured at 2.2 K, just above the helium $\lambda$ point, are shown in Fig. 9.7 along with the RF reflected power. Note the rapid temperature fluctuations with electron loading.

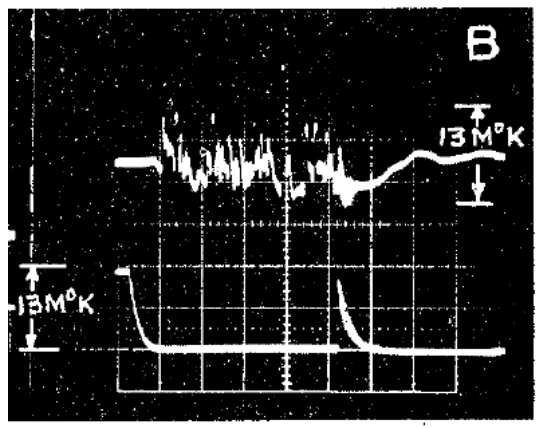

**Figure 9.7:** An oscilloscope screen shot showing typical temperature fluctuations (upper trace) and RF reflected power (lower trace) at 5.2 MV/m [29].

Using a program (MUFFIN) by Weingarten [30] to calculate the electromagnetic fields for a muffin-tin cavity, Padamsee found in 1979 the various multipacting levels caused by one-surface multipacting trajectories [31]. Fig. 9.8 shows trajectories for a 3$^{rd}$ order MP level. Thermometry (Section 9.7) showed MP levels at 3.8 and 5.2 MV/m for S-band muffin tin cavities. 2-cell S-band cavities, fired at 1900°C, were limited to fields between 5.5 and 6.6 MV/m (3$^{rd}$ order MP band), with one test reaching 9 MV/m (near the 2$^{nd}$ order MP band). 6-cell cavities were limited by quench at the inter-cell welds to about 2.2 – 2.7 MV/m (see Section 9.6 for quench limits and cures).

As discussed in Section 9.2, MP is a resonant process in which an electron avalanche builds up within a small region of the (quasi) pillbox shape $\beta = 1$ cavity (Fig. 9.6), due to a confluence of several circumstances. Electrons in the high magnetic field (equatorial) region of the cavity travel in quasi-circular and figure-eight like orbits returning to near the point of emission at about the same phase of the RF period as for their emission. The energy gain of 30 – 200 eV from the vertical component of the electric field is sufficient to generate secondary electrons on impact due to the secondary emission coefficient of the Nb surface, which is typically > 1 between 50 eV and 200 eV.

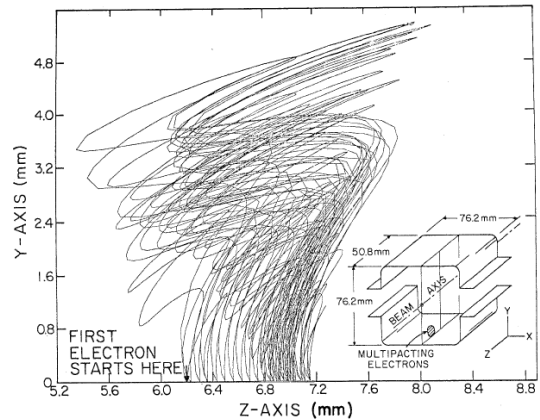

**Figure 9.8:** Calculated electron trajectories of the 3$^{rd}$ order MP in a muffin-tin cavity [31].

In 1978, Padamsee and Joshi measured the secondary emission coefficient of Nb between 200 and 2000 eV [32]. Nb samples were prepared by electropolishing and oxipolishing, techniques used for Nb cavity preparations at the time. The secondary coefficient of as prepared Nb was found to be between 1.4 and 1.6 with maximum located near 200 eV. Indeed, multipacting simulations had shown strong levels with electron energies near 200 eV, as Table 9.1 shows. The measurements were repeated after removing (by discharge cleaning) a few hundred Å from the surface to characterize clean Nb underneath, and then repeated once again after exposure to air, see Fig. 9.9. Later in 1986, Calder et al. [33] made similar measurements down to 100 eV to find coefficients as high as 2.8 for wet prepared Nb, reduced to 1.5 after baking at 300ºC and further reduced to 1.2 after argon discharge cleaning. These high secondary yields confirmed why 3$^{rd}$ order MP (with electron energies above 100 eV) was troublesome in SRF cavities and gave encouragement for MP reduction with baking and discharge cleaning.

## 9.5   *Multipacting cures*

### *Early efforts*

The first idea from Lyneis [28] to suppress or eliminate one-point MP was to modify the pill-box shape with a sharper corner radius that reduced the perpendicular component of the electric field at the outer wall. Calculations confirmed that the energy gain of MP electrons would fall to below 100 eV where the secondary emission coefficient of the Nb surface was expected to be less than one. A subsequent experiment on such a sharp-cornered anodized S-band niobium cavity confirmed absence of MP for fields up through third-order MP.

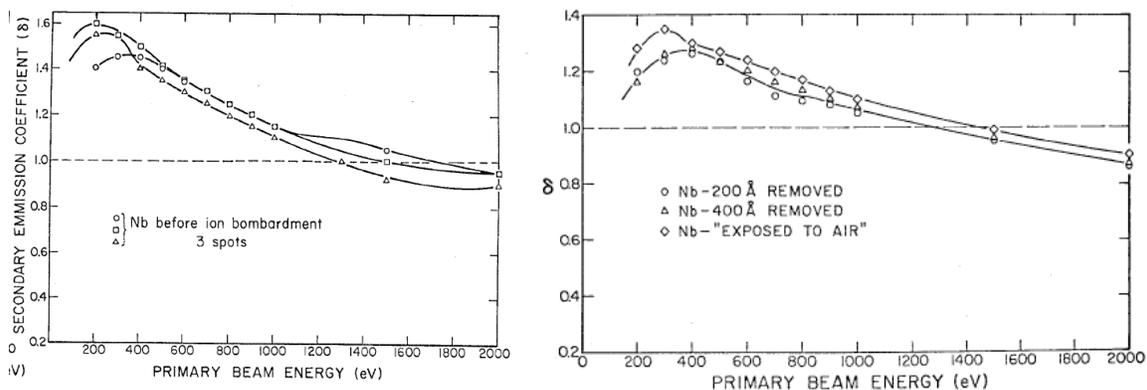

**Figure 9.9**: (a) Secondary electron emission for 3 spots of "wet prepared" Nb surface. (b) Secondary electron emission for "cleaned" Nb, and Nb "exposed to air" [32].

Kneisel and Padamsee followed the same idea in 1979 for the 2.86 GHz muffin-tin with some success [34]. When applied to a 2-cell, cavity, MP was suppressed in the π mode but remained active in the π/2 mode, as the cell shape was tailored to reduce the perpendicular component of the E field only in the important π mode. The 2-cell cavity was fabricated by machining from a solid niobium plate so that the desired profile could be made to close tolerances. Three tests were conducted on an unfired cavity. Test results were monitored with a thermometry system of 300 resistors (similar to the one discussed in Section 9.7). In the accelerating (π) mode there was no electronic activity discernable below 7.4 MeV/m accelerating field, as compared to the standard shaped S-band cavities which showed MP levels at 3.8 and 5.2 MV/m. Above this field some X-rays were observed out to 9 MV/m in some tests, but these could have been due to FE. However, in the non-accelerating (π/2) mode, where the perpendicular component of the electric field was significantly higher (4.5% of the axial field), multipactor-related effects were observed at low fields.

Around the same time, in 1980, Padamsee pursued a different way to reduce multipacting by incorporating grooves into the cup bottoms of the muffin-tin structure as shown in Fig. 9.10(a). Simulations showed grooves disrupt electron trajectories (Fig. 9.10(b)) in two distinct ways. If a returning electron hits the upper part of a groove, the secondary generated returns out of phase with the electric field. Deep inside the groove the electric field is substantially attenuated so that any secondary generated inside the groove does not gain enough energy to generate another electron. In 1981, several two-cell S-band cavities and one 2-cell L-band cavity were prepared with grooves using electro-discharge machining. With grooves, MP at $E_{acc}$ = 3 and 4 MV/m was eliminated in 2-cell S-band cavities to reach 8 MV/m. Grooving was able to eliminate MP in both π and π/2 modes, proving that grooving is superior to the tailoring the cavity shape to reduce the perpendicular component of E at the cup-bottom or equator.

By 1983, two full-scale 5-cell, 1500 MHz muffin-tin structures were built with HOM and fundamental power couplers and grooved cup bottoms. Both structures exceeded the high energy synchrotron design goal of 3 MV/m in vertical tests without limitations by multipacting, and with $Q$'s of 1.5 and $3\times10^9$. In 1983, Sundelin et al [35] successfully tested the first superconducting accelerating structures in a storage ring.

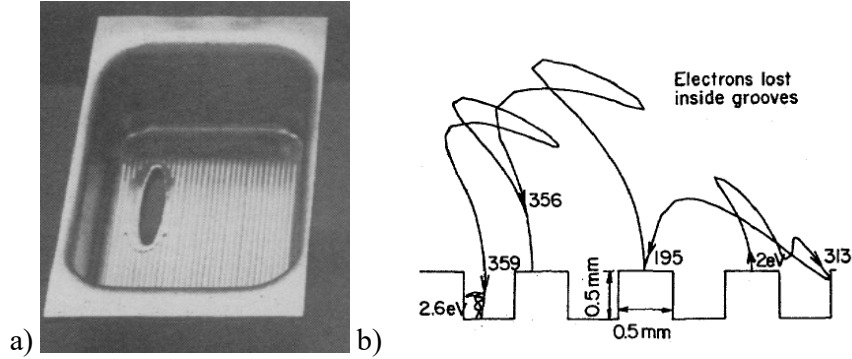

**Figure 9.10:** (a) One half of a muffin-tin cavity with grooved surface. The elliptical aperture is for coupling out higher order modes. (b) Electron trajectories disrupted by grooves.

*Decisive elimination of one-surface multipacting*

While efforts to understand MP were gaining strength along with attempts to cure MP, Parodi [36] at Genoa University serendipitously found high field performance in C-band (4.5 GHz) single cell cavities made by stamping half-cells from door-knob shaped dies, as they reported at the Applied Superconductivity Conference in 1979. Parodi chose this shape because he found doorknob forming dies in his workshop. These cavities reached a surprisingly high 26.5 MV/m gradient, with peak fields of 50 MV/m and remarkably high $Q$ levels of $2\times10^9$ below 20 MV/m. *The most extraordinary aspect of the reported performance was the absence of electron loading, as evidenced by the absence of x-rays.* Given that $3^{rd}$ order MP limited 1.3 GHz HEPL cavities at 2.15 MV/m, one would expect at 4.5 GHz similar $3^{rd}$ order MP at 7.4 MV/m, $2^{nd}$ order at 11.2 MV/m and $1^{st}$ order at 22 MV/m. For the Genoa cavities to reach 26 MV/m without any signs of "electron loading" was miraculous! It is true that $Q$ values started to fall above 20 – 24 MV/m in the three cavities Parodi studied, but the $Q$ fall off was gradual, and also without thermal breakdown. The absence of X-rays even when the $Q$ dropped also indicated lack of field emission. In retrospect, the $Q$-drop was most likely occurring due to the HFQS phenomenon, yet to be formally discovered (see Section 9.9).

*How did the doorknob cavities surpass all expected MP barriers?*

Returning from the ASC conference, Proch at Wuppertal University was inspired to determine why the door-knob shape cells showed no electron loading at the MP field levels expected. Using computer simulation, Proch and Klein [37] looked for one-surface MP in a cavity with a "spherical shaped" wall profile shown in Fig. 9.11(a). This was a watershed moment in the gradient history of SRF cavities.

Proch and Klein found that the electron trajectories were pushed by the electromagnetic fields of a spherical cavity to the plane of symmetry at the equator, where the vertical electric field component was nearly zero (exactly zero for perfect symmetry). The magnetic field varies along the cavity wall of the spherical cavity shape, so that there are no resonant electron trajectories, as secondary electrons travel to the equator in a few generations. Here the electric field is very low (nearly zero), so that secondary electrons cannot gain sufficient energy to multiply.

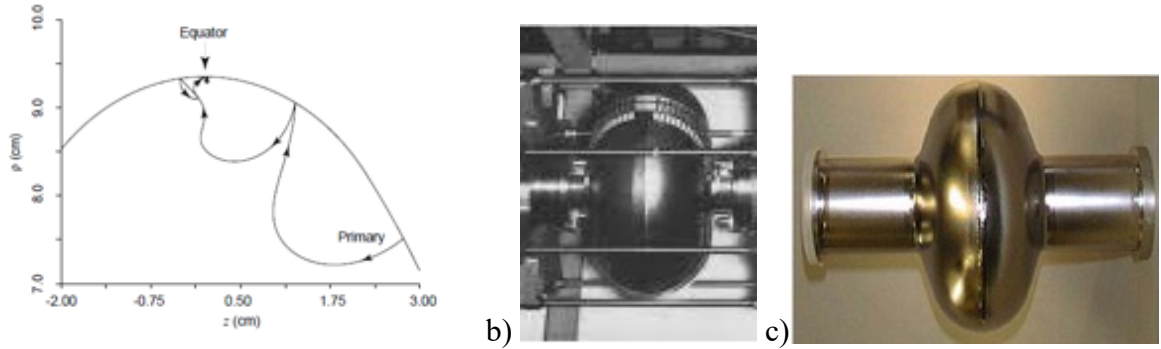

**Figure 9.11:** (a) Electron trajectories in a spherical cavity. Electrons drift to the equator where multipacting is not possible. (b) 500 MHz spherical cavity at CERN with thermometers. (c) 1500 MHz elliptical cavity at Cornell.

With the encouragement from Piel at Wuppertal, CERN – under Bernard and Lengeler – quickly adopted the spherical cavity shape to fabricate a single cell cavity (Fig. 9.11(b)) at the low frequency of 500 MHz [38]. (LEP upgrade was anticipated to need low RF frequency.) If one-surface MP were to present a problem, the cavity would be limited at 1.5 MV/m, judging from the strong MP at 3 MV/m in 1.3 GHz HEPL structures, and the simple frequency scaling law for MP. Instead, by the end of 1981, CERN reached 3 MV/m with high $Q$ at 4.2 K, and with no sign of one-surface MP. The cavity quenched between 4 – 5 MV/m. The spherical cavity shape had decisively won the day!

The same year (1981), Kneisel and Halbritter [39] published calculations and test results on a 3 GHz elliptically shaped cavity showing that the elliptical shape (Fig. 9.11(c)) is also free of MP. The basic principle is the same as for the spherical cavity. They invented the elliptical shape to provide higher mechanical strength than the spherical cavity which has straight walls. The sloped wall of the elliptical shape was also better for rinsing out acids during chemical treatment. The peak electric field was also lowered with an elliptical iris region. The 3 GHz cavity reached 11.5 MV/m accelerating on first power rise, which was nearly twice the expected multipacting level.

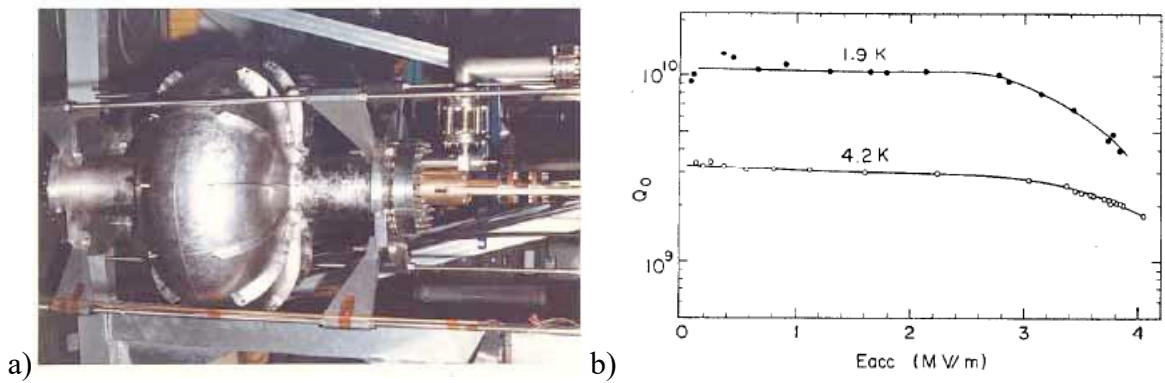

**Figure 9.12:** (a) KEK spherical cavity. (b) The cavity was a success: it reached the design gradient for TRISTAN of $E_{acc}$ = 3 MV/m at 4.2 K with a $Q$-value of $2.8 \times 10^9$ and very moderate electron loading.

Along with the 500 MHz spherical cavity result from CERN, one-surface multipacting was now conquered. Most labs adopted the elliptical shape rather than the spherical shape due to its better mechanical and rinsing properties.

Word of the successful spherical shape reached Kojima's group at KEK in Japan. At the end of 1979 efforts focused on the possibility of increasing the energy of TRISTAN from 30 GeV and higher by adding spherical shaped superconducting niobium cavities at 500 MHz to the existing normal conducting cavities. The spherical cavity was fabricated by MHI company and the electropolishing of the half cells took place at Nomura Plating. The KEK spherical cavity was a success [40]: it reached the design gradient for TRISTAN of $E_{acc}$ = 3 MV/m at 4.2 K with a $Q$ value $2.8 \times 10^9$ and very moderate electron loading (Fig. 9.12).

At Cornell, Kneisel in 1981 pushed to adopt the elliptical cavity and to put to bed the muffin-tin. The lower frequency of 1500 MHz provided a large enough aperture for synchrotron radiation to exit the cavity without impinging on the cavity wall. Five single cell and one 5-cell 1500 MHz niobium cavities of elliptical shape were fabricated and tested [41]. $Q$-values of 2 to $5 \times 10^9$ were obtained without high temperature firing. In single cells, an accelerating field of 7 MV/m were measured. In the 5-cell structure a gradient of 4.7 MV/m was achieved. By 1984, Kneisel [41] had tested several 5-cell elliptical cavities to reach gradients of 8 – 15 MV/m, clearly vindicating the superiority of the elliptical cavity. Kneisel prepared a series 15 tests of 8 single cell elliptical cavities from commercial niobium to obtain max fields of 35 mT ± 10 mT (8 MV/m).

In Nov. 1984, Sundelin et al. conducted a second storage ring beam test in CESR [42], this time using two 5-cell cavities of the elliptical cavity design, as well as higher purity Nb material (discussed in the next section) to achieve high gradients. In vertical tests the cavities reached gradients of 8, 9 and 15 MV/m. In CESR, the better of the two cavities reached 6.5 MV/m at a residual $Q$ of $5 \times 10^9$. This was the highest gradient reached in a storage ring test of a superconducting cavity. The cavities were operated successfully in CESR with 22 mA beam current.

The success of the storage ring test, along with the excellent vertical test results played a crucial role in 1985 for the subsequent selection of the Cornell 5-cell elliptical cavity design to build the CEBAF accelerator in Virginia [43]. With more than 300 cavities, the accelerator was commissioned as a 5-pass recirculating linac in 1995 to operate at 6.5 GeV [44]. The nuclear physicists' goal of a 2 GeV CW accelerator born at HEPL in the 1970's was finally realized, and even exceeded. Ultimately, CEBAF achieved an operating energy of 6.5 GeV by upgrading the performance of their cavities to 7.5 MV/m.

Returning to our discussion of the development and application of spherical cavities, by the summer of 1988 sixteen 5-cell spherically shaped cavities were installed in the TRISTAN tunnel [45]. The beam energy of TRISTAN was upgraded from 28.5 GeV to 30.7 GeV. A second set of 16 cavities were installed at the beginning of August 1989. In the fall of 1989, a beam energy of 32 GeV was achieved [45].

The majority of the 32 cavities in TRISTAN reached a gradient of 10 MV/m in the vertical tests. The subsequent assembly into pairs with input couplers, HOM couplers and tuners was done in a class 100 clean room; but some assembly steps needed to be done in a less clean environment which resulted in a 30% degradation of the operating gradient.

The upgrade of TRISTAN by 1989 was the first large-scale successful demonstration of SRF technology in an accelerator and was truly a pioneering effort due to a visionary leadership by Kojima of a dedicated and immensely competent group at KEK.

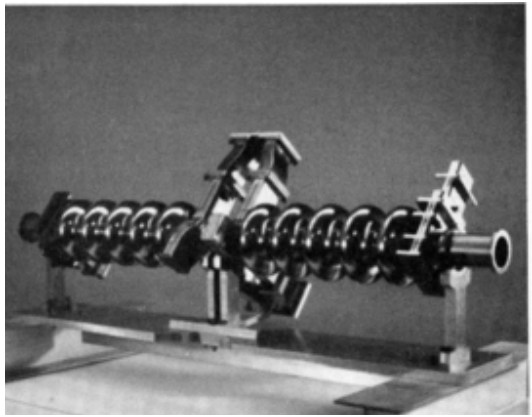

**Figure 9.13:** CEBAF two-cavity string of the Cornell 5-cell 1500 MHz SRF cavities.

*Two surface multipacting*

At CERN Weingarten in 1983 [46] discovered in experiments with 500 MHz cavities that a weak MP barrier still remained in the spherical cavity. This was identified by his calculations to be two-point MP (Fig. 9.14). Multipacting conditions exist when electrons travel to the opposite surface in half an RF period (or in odd-integer multiples of half an RF period). Two-point MP survives near the equator of the elliptical cavity because the electron energies remain between 50 – 100 eV, near the unity cross over of secondary yield. However, the barriers are easily processed since the electron energy was generally < 70 eV where the secondary emission yield is just near unity. Two-point MP still rears its head for the high gradient elliptical cavities in European XFEL, LCLS-II and other cases. These cavities show MP at accelerating fields of about 17 to 21 MV/m [47]. A general solution to two-point MP is still desirable in the community. Perhaps incorporating a few grooves in the largest diameter region of the equator would solve the problem.

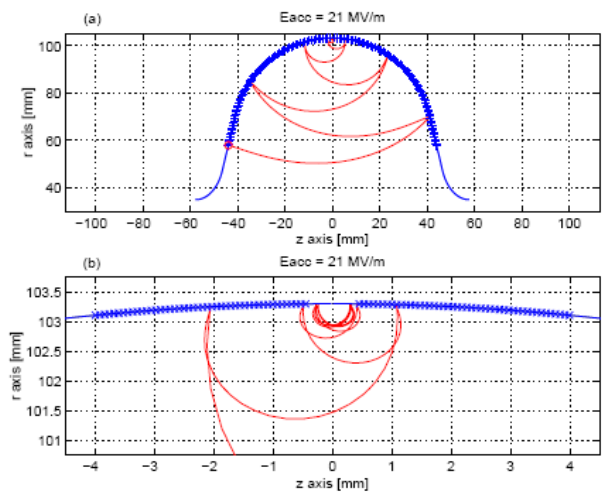

**Figure 9.14:** Two-point multipacting in a spherical cavity.

## 9.6  Thermal breakdown, quench of superconductivity

### Understanding of thermal breakdown

With multipacting finally overcome, quench of superconductivity became the next limiting mechanism, at about 5 – 6 MV/m. Before 1980, there were two theories about the root cause. One was that quench was due to a "thermal breakdown" which originates at sub-millimeter-size regions that have RF losses substantially higher than the surface resistance of an ideal superconductor. These regions earned the name "defects". Another interpretation [48] of quench was that the breakdown of superconductivity originates when the RF magnetic field exceeds a local critical magnetic field causing an abrupt local phase transition from the superconducting to the normal state. The local RF critical field was presumed to be depressed due to the presence of the defective region.

This magnetic transition interpretation was definitively ruled out by a special experiment invented by Proch in 1980 [49]. In this experiment the local RF magnetic field at the breakdown spot was increased by superposing the field of a second cavity mode. A 2-cell cavity was first excited separately in the two modes (called $\pi$ and $\pi/2$) of the fundamental passband, and the breakdown field for each mode was measured separately. The same breakdown spot for both modes was observed by thermometry. In the next stage, both $\pi$ and $\pi/2$ modes were simultaneously excited, and different ratios of field amplitudes were adjusted to obtain breakdown at the same spot. If the magnetic instability model was applicable, one would expect that breakdown would take place at a certain well-defined magnetic field, so that

$$H_\pi + H_{\pi/2} = \text{constant}.$$

If a temperature instability was responsible for breakdown, one would get

$$H_\pi^2 + H_{\pi/2}^2 = \text{constant}.$$

In several tests cavities for which the quench field ranged from 15 to 50 mT, the second result was unambiguously obtained, as shown in Fig. 9.15. These experiments definitively showed that the breakdown level depends not on the local $H$, but on the local $H^2$, or power. Furthermore, when equal power was applied in both modes, the surface magnetic field at the breakdown location was found to be $\sqrt{2}$ times higher than the quench field in either mode alone, definitely ruling out the magnetic instability mechanism. Proch cleared up the ambiguity about the magnetic or thermal character of breakdown.

More evidence that quench is of thermal origin came from thermal maps that detected ohmic heating at the defect, and tracked the temperature rise near the defect all the way from low fields to the quench (see temperature map in Section 9.7). Breakdowns in several cavity tests were caused solely by thermal runaway, i.e. no magnetic transition was involved. In these tests the surface magnetic field at the defects ranged from 13 to 97.5 mT. The temperature excursions outside the defects were usually found to increase as $H^2$.

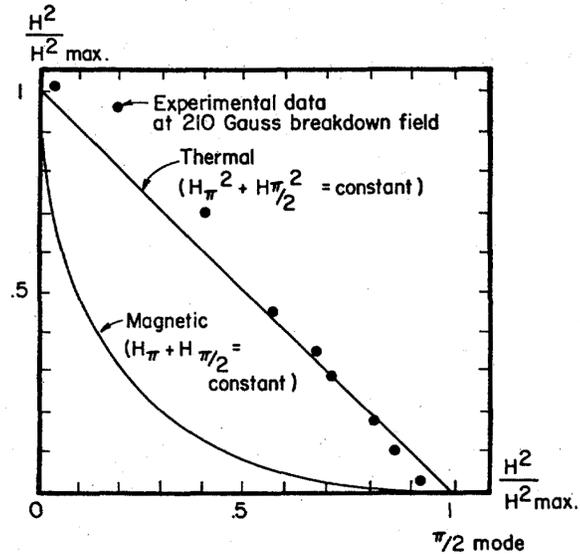

**Figure 9.15:** Experimental data unambiguously showing that the quench is due to thermal breakdown, and not initiated by a magnetic transition.

*Identification of defects by thermometry and cavity dissection*

In 1983, Weingarten and Padamsee [50] at CERN first located quench producing defects (Fig. 9.16) in 3 GHz cavities with thermometry, and then dissected the cavity to analyze the defects in an SEM. This was the first time that a cavity was sacrificed for valuable information. Similar efforts since that time have found that typical defects are chemical stains, foreign metal inclusions, pits with sharp edges, metal burrs from scratches, voids, weld beads and other types of welding mistakes [51-54].

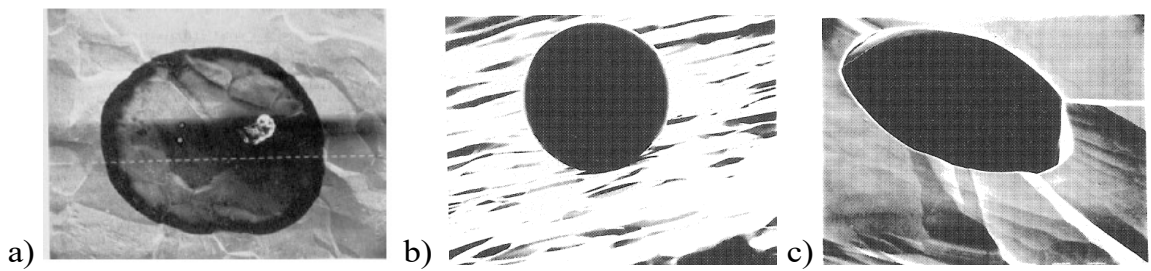

**Figure 9.16:** Defects located by thermometry and analyzed in SEM after dissection [50]: (a) Chemical residue, diameter 400 microns, quench field 3.4 MV/m. (b) Nb sphere from a weld bead, diameter 80 microns, quench field 6.8 MV/m. (c) Microscopic hole in a weld, quench field 8 MV/m.

*Numerical model calculations*

In 1979, Deniz and Padamsee [55] developed a simple numerical model calculation to illustrate the essential features of thermal breakdown initiated by small normal-conducting defects. The code was based on an iterative solution of heat flow equations to solve for the equilibrium temperature in the vicinity of a small circular defect. Using this code, they studied the dependence of the breakdown field on defect size, defect resistance, bath temperature, RF frequency, temperature dependent thermal conductivity, Kaptiza conductance, residual resistance and heat transfer limits to the He bath. For the first time it became possible to sort out in detail the influence of the many variables. A main result from the model studies was that the thermal breakdown field $H_{max}$ follows the simple law [50]:

$$H_{\max} = \sqrt{\frac{4\kappa(T_c - T_b)}{aR_n}}. \tag{9.6.1}$$

where $a$ is the radius of defect, $R_n$ is the resistance of defect, $\kappa$ is the average thermal conductivity, $T_c$ is the transition temperature, and $T_b$ is the bath temperature.

The model calculations revealed the process of thermal breakdown. Defects absorb power from the microwave fields, heat up the neighboring superconductor (as shown in Fig. 9.17) and eventually drive it above the critical temperature (9.2 K), leading to the thermal breakdown instability. Equation (9.6.1) showed that a natural countermeasure was to increase the thermal conductivity of the cavity wall so that higher dissipation at many naturally occurring imperfections on the surface of Nb cavities could be tolerated. Then defects would be able to tolerate more power before driving the neighboring superconductor into the normal state. As we discuss later in this section, thermal conductivity of Nb depends strongly on the purity which can be characterized by RRR.

$$\text{RRR} = \frac{\text{resistivity 300 K}}{\text{residual resistivity at low temperature (normal state)}}$$

At 4.2 K, the thermal conductivity of niobium is given approximately by a simple relation:

$$\kappa = 0.25 \text{ (W/m·K)} \cdot \text{RRR}.$$

Therefore, one can gauge the thermal conductivity of niobium with the purity of niobium by measuring the RRR. Because of this relationship, RRR and $\kappa$ are often used as equivalent quantities. It has become customary to quote the RRR as a convenient gauge of the total impurity content of Nb.

Computer simulations predicted the performance benefits for niobium cavities with improved niobium thermal conductivity, discussed further below. Fig. 9.18(a) shows from model predictions how the maximum surface RF magnetic field increases roughly as the $\sim\sqrt{\text{RRR}}$ [56], which is equivalent to $\sim\sqrt{\kappa}$.

The thermal model was also used to calculate maximum fields possible for defect free cases. For zero residual resistance, the low-frequency maximum field was found to be higher than $H_c$ at 1.5 K, so that thermal instability should not be a problem at low frequencies. At 6 GHz the maximum field was found to be less than 170 mT, decreasing to 110 mT at 8.6 GHz. These values are in good agreement with earlier results of defect-free model calculations of Lyneis, as discussed below.

In 1974, Hillenbrand et al. [57] made the first model calculation of the instability threshold for niobium with the several oversimplifying assumptions of no defects. In 1976, Lyneis et al. [58] calculated thermal instability thresholds for Nb and NbTa using a line defect model chosen because of the mathematical simplicity. Lyneis also made calculations for the defect-free case.

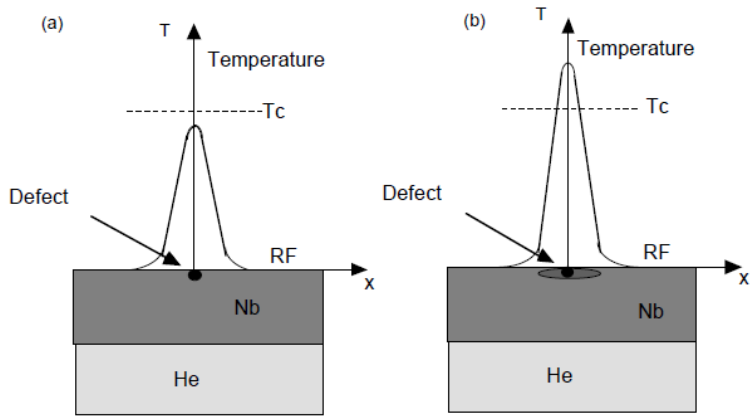

**Figure 9.17:** Thermal breakdown: (a) Calculated temperature rise in the vicinity of a defect. (b) When the temperature of the superconducting niobium just outside the defect rises above $T_c$, the defect grows in size unstably, and so does the power deposited. This is called the quench [1].

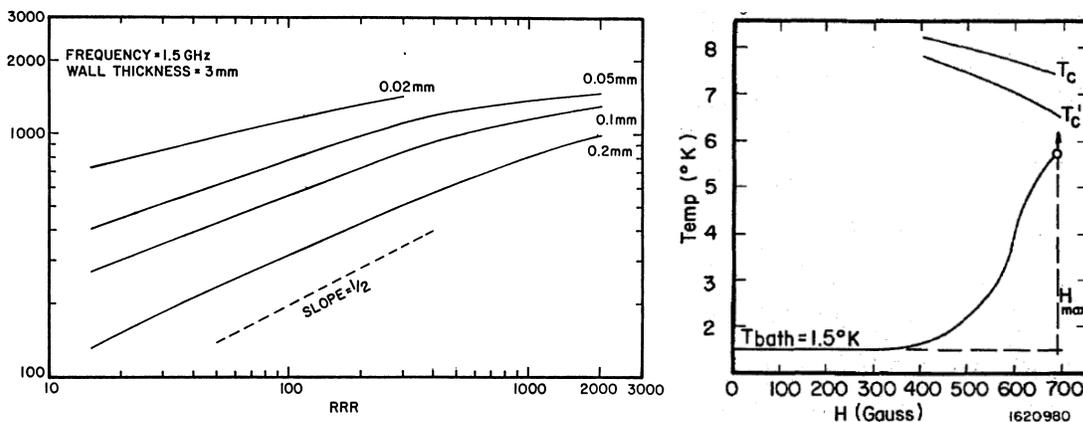

**Figure 9.18:** (a) Thermal model predictions for the breakdown field of various normal conducting defects with radii given for each curve. For comparison, the quench field calculated from equation (9.6.1) is shown as the line with slope 1/2. Note that the full simulations show similar slopes $\sim\sqrt{RRR}$ but higher field values because they include the temperature dependent thermal conductivity, the average of which is higher than the thermal conductivity at 4.2 K used to plot the result from equation (9.6.1). For these cavities $H_{pk}/E_{acc} = 47$ Oe/MV/m. (b) A defect that becomes unstable at 72 mT is chosen. The lower curve shows the calculated temperature of the RF surface in the immediate neighborhood of the defect as the rf field level increases. At the highest field for which a stable solution exists the temperature near the defect is 1.2 K below the "critical" temperature $T_c$. Above this field the temperature increases unstably (as indicated by the arrow) so that a thermal explosion to $T_c$ occurs without magnetic transition.

Later, in 1992, Röth et al. [59] at Wuppertal developed a variable mesh density thermal model code for reliable calculations with the smallest defects, down to 1 micron in diameter. Fig. 9.19(a) shows the results for breakdown fields for a large range of defect sizes and RRR values. Fig. 9.19(b) shows the temperature rise near the defect versus field for various defect sizes.

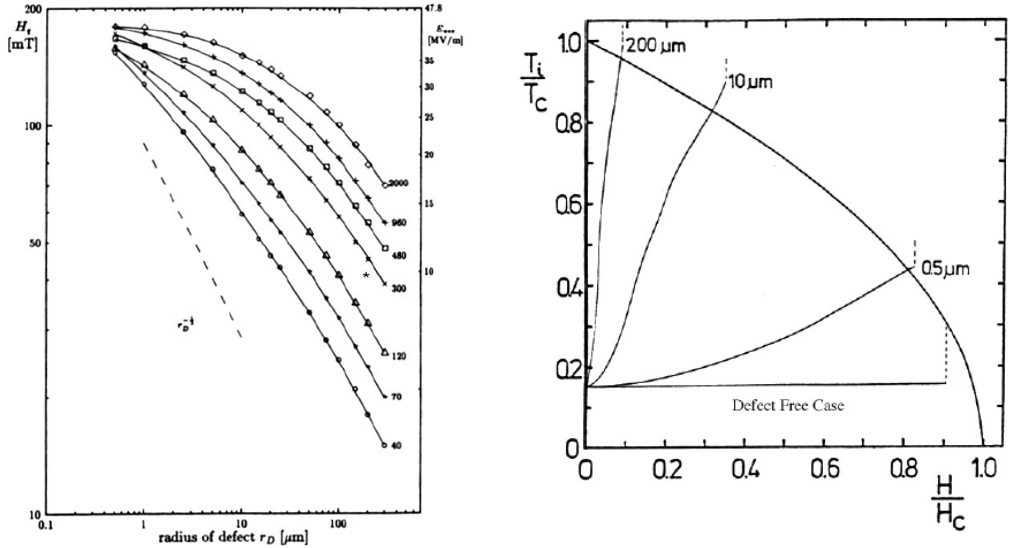

**Figure 9.19:** (a) Calculated thermal breakdown field vs. defect size for niobium of various RRR values, shown next to each curve. (b) Temperature vs. RF field strength for the region just outside the defect. The calculations are for various defect sizes, including a case with no defect. RRR 30 niobium was selected for this case [59].

## *Thermal conductivity of niobium and niobium purity*

In 1980 Krafft [60], a graduate student at Cornell, compiled a review of the various measurements of the thermal conductivity of Nb and discussed the various factors that play a role in the physics of the important parameter. Foremost among these factors is the RRR, the residual resistance ratio. The higher the RRR, the higher the thermal conductivity. An effective way to monitor the purity (and so the thermal conductivity) is to keep track of the residual resistivity ratio (RRR) of the Nb. This important Nb specification is the ratio of the DC electrical resistance of Nb at room temperature to the DC resistance of Nb at 4 K but in the normal-conducting state, when the resistance is mostly due to the impurities (residual resistance). The RRR for 1 ppm of each of the impurities is well documented [61, 62].

The physics of the temperature dependence of Nb thermal conductivity is very interesting. Electrons and phonons (quantized lattice vibrations) are the major heat carriers in metals. As niobium is a superconductor, the thermal conductivity of niobium drops precipitously below $T_c$, as electrons condense into Cooper pairs at an exponentially increasing number with decreasing temperature. Cooper pairs are not scattered by the lattice vibrations, and therefore cannot conduct heat. Between $T_c$ and 4 K, a significant, though a small, fraction of electrons is not yet frozen into Cooper pairs and so can still carry heat, provided that the electron impurity scattering for the non-condensed electrons is low. As the 1981 review of measurements by Schulze [61] at the Max Planck Institute states, the most significant electron scattering impurities are the interstitial ones, namely O, N, C, and H. Schulze used ultra-high vacuum ($10^{-9}$ torr) degassing at temperatures close to the melting point to produce Nb with RRR as high as 30,000, near the ideal value with zero impurities!

Another dominant impurity is Ta, which comes from the starting Nb ore, but this impurity does not pose as much harm to thermal conductivity because Ta atoms are substitutional, not interstitial, so that Ta does not scatter electrons much. For example, one ppm of oxygen (a dominant impurity) will by itself limit Nb RRR to 5000. By comparison, it will take 500 ppm of Ta to give a similar RRR.

As mentioned earlier, one can gauge the thermal conductivity of niobium with the purity of niobium by measuring the RRR which requires a measurement of the low-temperature resistivity of niobium in the normal state. The interstitial impurities have an equivalent effect on the low temperature electrical and thermal conductivity.

If there were no impurities in Nb at all, the ideal RRR of niobium is 35,000 and due only to electron-phonon scattering. Phonons begin to play a bigger role as heat carriers below 4 K. As electrons condense into Cooper pairs, electron–phonon scattering decreases. Below about 4 K, the thermal conductivity from phonons rises, as electron-phonon scattering falls, leading to the phonon peak near 2 K. With decreasing temperature, the number of phonons decreases proportionately to $T^3$. Ultimately, the value of the phonon conductivity maximum is limited by phonon scattering from lattice imperfections, such as the density of grain boundaries. If the crystal grains of niobium are large (cm scale), because of annealing at high temperature, or due to slicing sheets from the melted Nb ingot, one observes a large phonon peak. *However, the phonon peak does not play a role in stabilizing the heat produced at defects, because the temperature of the niobium surrounding the defect rises with field.* Thermal model calculations discussed above show that the 4 K to 9.2 K average thermal conductivity is important in determining whether the temperature outside the defect will cross $T_c$ and cause thermal breakdown.

Thermal conductivity functions used in the computer model simulations discussed in the previous section are shown in Fig. 9.20. In 1996 Koechlin and Bonin [63] developed a very useful parametrization of Nb thermal conductivity from just the RRR and the grain size of the Nb. Fig. 9.21 shows their results.

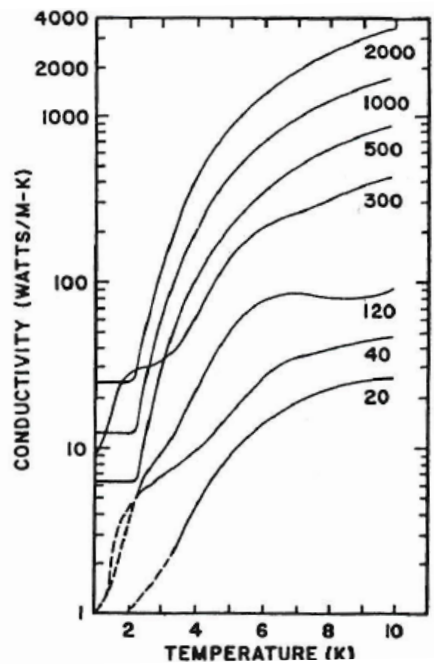

**Figure 9.20:** Dependence of Nb thermal conductivity on temperature for different RRR [64]. The temperature near the defect rides between the bath temperature (2 – 4 K) and $T_c$.

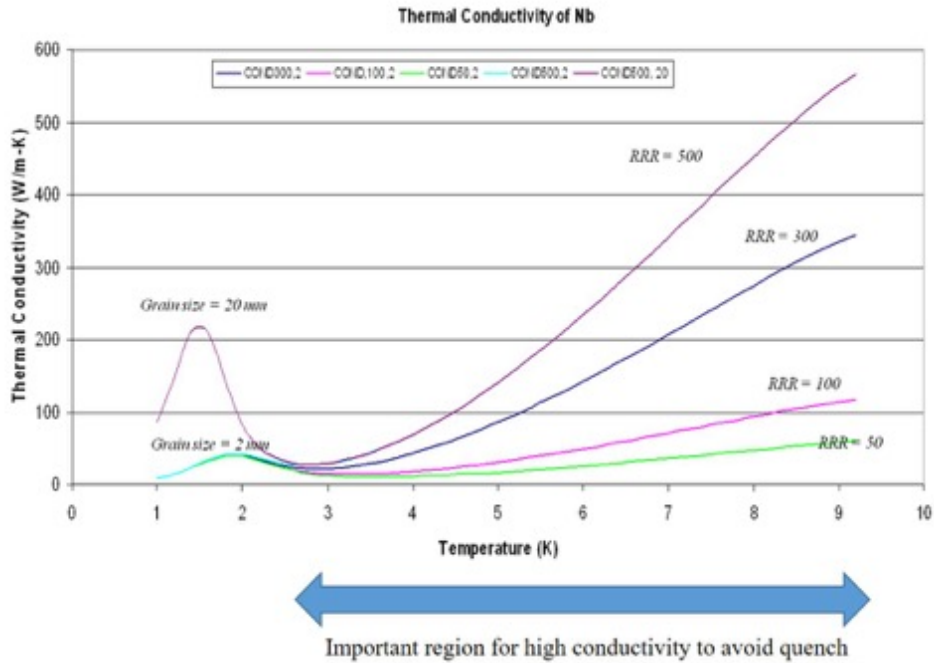

**Figure 9.21:** Results from a very useful parametrization [63] of Nb thermal conductivity from the RRR and the grain size of the Nb.

*Cures for thermal breakdown*

An obvious approach to avoid quench is to use great care in preparing the niobium material and arranging the fabrication procedures and electron beam welds without introducing any defects. This is generally good and important practice to follow as much as practical but becomes a tall order on a large scale. To reduce the chance of getting defects, the starting niobium sheet can also be screened by the eddy-current scanning method, discussed later [65, 52]. On a large scale, it becomes impossible to ensure that there will be absolutely no defects, especially in large area cavities, or when dealing with hundreds, or even thousands of cavities.

The best insurance against thermal breakdown is to raise the thermal conductivity of the niobium especially from the bath temperature (2 – 4 K) to $T_c$. With the higher thermal conductivity, a given defect can tolerate higher rf dissipation, the temperature of the Nb near the defect can rise safely with the applied field without crossing $T_c$, and higher gradients become possible.

*Methods to improve niobium purity*

We now turn to historical development of measures to improve the thermal conductivity of niobium, and thus to overcome thermal breakdown. Three methods have been pursued: In chronological order the methods are: 1) UHV degassing of Nb near 2000ºC; 2) Solid state gettering; 3) Improving conditions for electron beam melting in commercial Nb production.

The first method, as shown by Schulze [61, 62], involves heating niobium in an excellent vacuum ($10^{-9}$ torr) at $T > 1900°C$ and for long periods (5 to 10 hours). Since niobium has a very high affinity for oxygen, very high temperatures are necessary to degas oxygen out of Nb into the vacuum. Even at 1900°C the removal of interstitials takes place very slowly.

Following this method, Padamsee in 1984 [66] purified a series of 8.6-GHz niobium cavities (1" diameter) by degassing at 2000°C using resistive and induction heating. RRR values up to 1200 – 1400 were obtained. To exclude possible benefits that may arise from an improved surface after heat treatment, the cavities were chemically etched after purification so that the same final surface preparation was present, independent of the bulk RRR. The RF test results from these cavities provided the first clear proof-of-principle that improving niobium purity (RRR) leads to higher quench fields, as shown in Fig. 9.22. Here, the measured quench fields are compared with thermal model calculations (discussed above) for various defect sizes and RRR values.

Although valuable as a research technique, the high temperature outgassing method is not practical for accelerator structures due the need for ultrahigh vacuum and high temperatures, which leads to severe deformations, creep, and loss of yield strength of Nb.

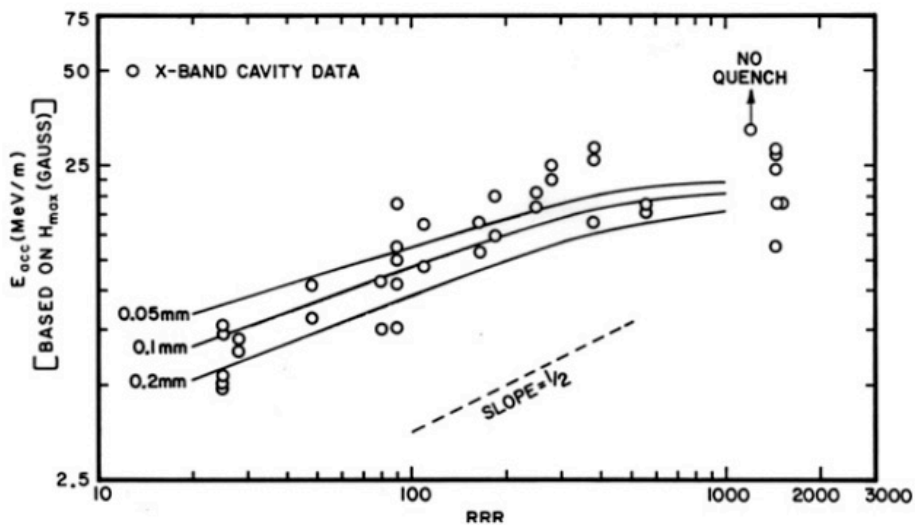

**Figure 9.22:** Measured quench fields of 8.6-GHz cavities after varying the niobium RRR by outgassing at high temperature. The solid lines are calculations for various defect sizes. $H_{pk}/E_{acc}$ = 47 Oe/(MV/m).

*Solid state gettering*

In the same year 1984, Padamsee [64, 67] pursued improving niobium purity via solid state gettering, applied for the first time to raise the RRR of niobium cavities by using yttrium to coat the niobium surface. Yttrium has a higher affinity for oxygen than does Nb [68-70]. The coated niobium is heated to a temperature > 1200°C – 1250°C in a vacuum of $10^{-5}$ torr so that oxygen diffuses rapidly to the surface. The treatment lasted about 4 hours. At 1250°C, the vapor pressure of Y is near $2\times10^{-5}$ torr so that several microns of Y were deposited on the cavity surface. The mobile interstitial impurity atoms inside Nb sink into the foreign metal layer (yttrium) when they arrive at the surface of the niobium. During the treatment, the Y foil wrapping also intercepted the impurity atoms from the furnace, preventing the furnace vacuum from contaminating the Nb cavity. Therefore, the purification procedure was successfully carried out in a diffusion pumped furnace with a vacuum of $10^{-6}$ torr.

The coating, purification and protection operations were combined into one step by the vapor pressure of yttrium, large enough to form an evaporated layer at the diffusion temperature. Direct contact between the cavity surface and the yttrium foil was avoided by interleaving the yttrium sheets with perforated Nb sheets. With solid state gettering for 4 hours at 1200ºC, a factor of 3 - 4 improvement in RRR became possible for commercially available niobium (RRR = 20 – 30), corresponding to nearly complete removal of the oxygen, the dominant impurity. After the purification stage, the getter material and the underlying compound layer were chemically etched away. Commercial niobium of RRR = 30 was improved to RRR = 90 to 120. By applying the solid state gettering to process multiple e-beam melted Nb by Fansteel (see next section) the RRR was improved from 60 – 80 to 310 - 360. A sample of high purity Nb from Ames Laboratory with RRR = 165 was improved to 540. For cavity fabrication, the post purification technique was applied after the half-cell forming stage, because the grain growth at high temperature would have destroyed the mechanical workability of the niobium at the sheet stage, making it impossible to form half-cells by stamping.

Many tests were carried out with 1.5 GHz elliptical cavities of RRR = 80 obtained by yttrification of reactor grade Nb RRR about 25. Most tests were limited by FE to below 65 mT (which translates to 15.5 MV/m in the TESLA elliptical cavity structure). Three tests reached quench fields with an average field of 66 ± 10 mT. These results were compared with many tests on elliptical cavities made from commercial Nb of RRR 25 (see Section 9.5) which reached quench field average of 35 ± 10 mT (which in turn translates to accelerating field of 8.3 MV/m) [41]. The gain in gradients (×1.9) is very close to the $\sqrt{RRR}$ (1.8) expected from simple analytical and computer models.

Subsequently, two 5-cell elliptical cavities were fabricated from yttrium purified reactor grade Nb. In multiple tests, one cavity reached 7 and 8 MV/m and the other reached 10 and 15 MV/m after a couple of repair cycles (see Section 9.5). These two 5-cell cavities were used for the CESR storage ring beam test by Sundelin et al., where a maximum accelerating field of 6.5 MV/m was reached in one structure [42]. The second cavity reached 2.4 MV/m during the beam test. Later the cause of this limitation was identified as a dirt spot in the center cell and was removed after the beam test by rinsing with detergent, water and methanol, but no acids. After the rinsing, the second cavity reached 11 MV/m at a $Q$ value of $6\times10^9$ and 12 MV/m at a $Q$ value of $2\times10^9$, when substantial field emission loading, but no breakdown was encountered.

Soon after the invention of solid state gettering by yttrium, Kneisel in 1986 used titanium as effective solid-state getter [71]. Titanium is less expensive than yttrium. However, the vapor pressure of titanium is lower than for yttrium, so that higher temperature or longer times are needed. For example, to remove oxygen in a few hours, titanium must be used at 1250 – 1300ºC for 10 hours. Since titanium also diffuses into niobium to a substantial depth (100 μm) along the grain boundaries [72], a heavier (than for Y) chemical etching was necessary after the post-purification step. The outside surface of a cavity must also be etched to reestablish a good Kapitza conductance [73]. Titanium does have the intrinsic capability to also remove nitrogen and carbon by solid state gettering due to its appreciable affinity to these impurities [74]. But the diffusion rates of nitrogen and carbon are much lower than for oxygen. Therefore, very long gettering times are necessary for removal of those impurities. Using titanium for more than 50 hours, RRR values > 1000 have been achieved in samples of starting RRR = 200 [75].

Kneisel purified two single-cells and two five-cell elliptical cavities to reach RRR values > 300, starting from RRR values of 160 and 45. The accelerating fields for the single-cells improved from 8.5 to 9.8 MV/m, and from 6.1 to 13.6 MV/m and the 5-cell improved from 7.9 to 11.9 MV/m. Niobium cavities with higher RRR by solid state gettering with yttrium or titanium proved the effectiveness of high purity, high thermal conductivity for mitigating quench.

*Electron beam melting at Nb producing industry*

The most effective approach to increase the thermal conductivity of niobium was to remove the interstitial impurities by improving electron beam melting practices used for refining the ingot at the Nb industry. This required multiple melting cycles in a moderately good vacuum ($< 10^{-5}$ torr) in the electron-beam furnace at the melting temperature of niobium (2470ºC). It was also important to melt the ingot slowly in order to achieve equilibrium [74].

Before 1984, RRR of commercial grade Nb sheets available from Nb suppliers measured between 20 and 30. Typical thermal breakdown field for niobium cavities was about 20–30 mT. The corresponding $E_{acc}$ ($\beta = 1$ cavities) was 6 to 7 MV/m for 1500-MHz and 3000-MHz single cells, and 5 MV/m for larger area, 500-MHz single-cell cavities. The best gradient for the low-RRR cavities was 10 MV/m, and 8 MV/m for 500-MHz single cells. Large area cavities usually quench at lower field due to the higher probability of encountering defects. A comparison with the simulations of Fig. 9.22 suggest that the typical normal conducting defect radius is 100–200 µm.

In an early effort (1984 – 1985), Padamsee collaborated with Fansteel [74] as well as with Ames laboratory to produce high RRR Nb by improved ingot melting practice. Fig. 9.23(a) shows the calculated decrease for O and N concentrations with number of melts assuming a typical ingot size, melt rate and partial pressures ($2\times10^{-5}$ torr) in the furnace vacuum. Starting concentrations are typical of first melt after consolidation of the Nb material to prepare the starting ingot. The calculation shows that 3 to 4 melts are sufficient to reach the equilibrium impurity concentrations needed for RRR = 100 Nb. To achieve higher purity requires further improvement in furnace vacuum, not necessarily more melt cycles. Fansteel was successful in supplying RRR = 80 – 90. Subsequently, Padamsee worked with Wah Chang to provide even higher RRR [74].

Hereaus in Germany quickly followed, with encouragement from Piel and Mueller in Wuppertal and from Weingarten at CERN. Between 1984 and 1987, over 8 tons of Nb ingot and sheet with RRR up to 200 were produced by Wah Chang, Heraeus, KBI, and Tokyo Denkai. Fig. 9.23(b) shows the rapid improvement of RRR over the following years. The quality of high RRR ingot was preserved when converting to the final sheet product by selecting industrial annealing furnaces with a good vacuum ($10^{-5}$ torr) and by incorporating protective measures such as wrapping Nb sheets with Ti foils.

Niobium is now available with RRR = 250 to 400 from U.S., European, Japanese and Chinese suppliers who use the techniques of multiple and slow melting. Major suppliers are now able to provide sheet Nb with RRR > 300, yield strength of 14,000 psi and grains size ASTM 6-7. For a short time, Russian niobium was available with RRR = 500 to 700 [76, 77] from Giredmet.

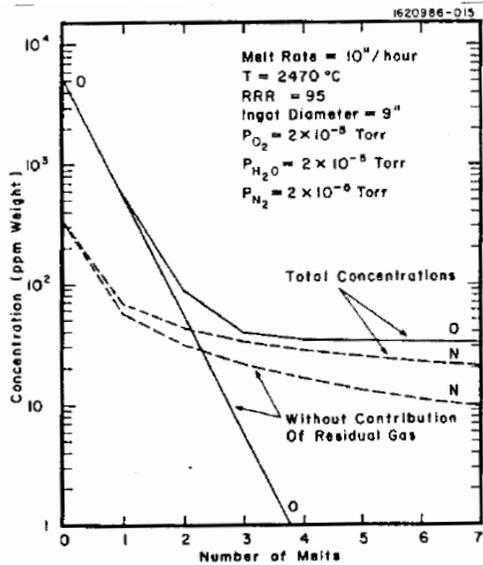 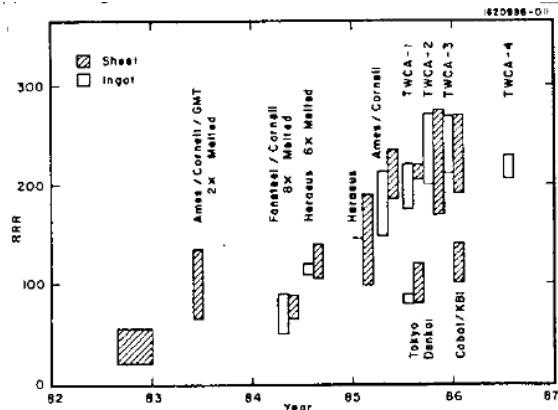

**Figure 9.23:** (a) Calculated degassing rates for oxygen and nitrogen by electron beam melting Nb ingots. Starting concentrations are typical of first melt after consolidation. (b) Progress in Nb Ingot and sheet purity. Shaded areas represent sheet data [74].

In 1986, Saito at KEK [78] worked with Tokyo Denkai, a Japanese Nb supplier, to improve the RRR of the Nb ingot to 110. Ingots with RRR = 80 – 110 were made by melting 4 – 5 times. The vacuum pressure of the industrial EBW system used saturated at $10^{-4}$ torr after the third melting. Titanium foil protection was used to limit for RRR loss during annealing of sheets and also for hydrogen degassing of cavities after EP in the furnace vacuum of $10^{-5}$ torr. They could increase the RRR of the niobium material to over 160. The achievable accelerating fields of single cell cavities were increased from 5 to 12 MV/m, again proportional to $\sqrt{RRR}$ as expected from the thermal model (Fig. 9.24(a)). In late 1990's [65] DESY confirmed a similar correlation for 9-cell TESLA cavities prepared by BCP.

KEK made 5-cell structures at 500 MHz from Nb with RRR values from 115 to 170 [78]. A distribution of accelerating fields achieved in vertical tests (Fig. 9.24(b)) shows a mean value of about 10 MV/m, a great achievement for a large-scale system. Unfortunately, the introduction of couplers and final assembly contaminated the cavities to result in a final operating field of 5 MV/m. In 1989, the KEK 500 MHz SRF system became the largest SRF system, providing more than 240 million volts [79]. The previous record was held by the Stanford University HEPL SCA which operated 34 meters of Nb cavities, built in the 1970s, providing a total of 50 million volts [17, 80].

Nb industries continued to be very helpful in the SRF quest for higher gradients. During the 1980s and 1990s RRR of Nb improved by an order of magnitude (from about 30 - 40 to 300 - 400) with the cooperation of electron-beam melting industry [74]. This means that interstitial impurity levels fell from about 150 ppm to < 15 ppm total. As a result, cavity gradients made from the purer, higher RRR Nb rose on average by a factor of three before the next limiting mechanism (field emission) kicked in. The correlation between higher RRR and increased quench field is clear from the data of single cell [66, 74, 78] and later nine-cell cavities made of fine-grain Nb and treated by BCP [65].

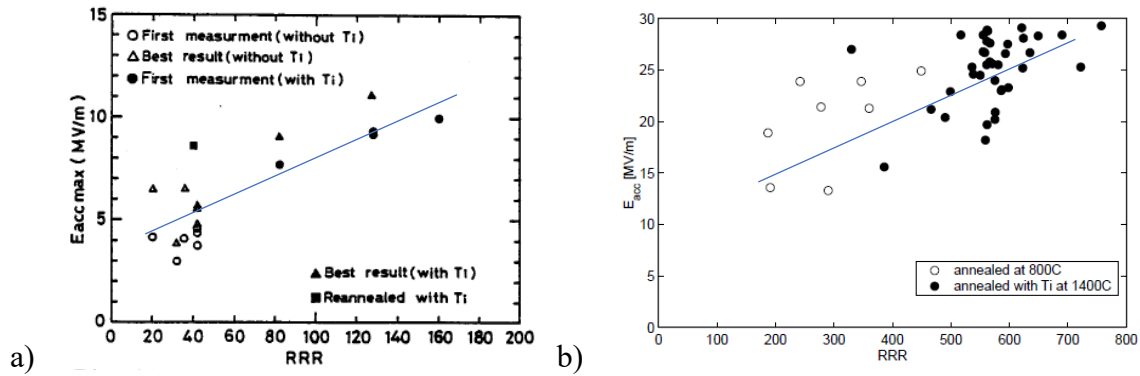

**Figure 9.24:** (a) Correlation between RRR and achievable maximum field gradients for (a) 500 MHz single cell cavities at KEK. (b) 1300 MHz cavities at DESY [65].

*Other solutions to thermal breakdown: Improved electron beam welding*

A common breakdown cause for the 2-cell and 6-cell muffin tin cavities made by the methods of electron beam welding stamped half-cells (and other components) was found by thermometry to be at the electron beam weld between cells. After mechanically polishing the weld beads, several 1-cell, 2-cell and 6-cell S-band muffin-tin cavities achieved $Q$ 's between $3\times10^9$ and $6\times10^9$ and gradients up to 2.7 MeV/m. After firing at 1900°C in the Brookhaven Lab UHV furnace, $Q$ values between $6\times10^9$ and $14\times10^9$ and gradients up to 6.9 MeV/m were reached. Comparable $Q$ values and significantly higher field values were obtained in the 1-cell and 2-cell cases.

The UHV firing improved the defective weld areas in the 1.5 GHz muffin-tin multicell structures, to some extent. Sundelin proposed [42] an interesting cause for the weld defects. A focused electron beam produces a vapor column in the metal while welding, and the weld puddle solidifies with vacuum bubbles still present. Bubbles immediately below the cavity surface impact cavity performance by interfering with heat transport. If a bubble opens up by chemical processing its sharp edges enhance local fields.

Sundelin and Kirchgessner [21] solved the focused beam welding problem using the rhombic raster welding technique. The electron beam used for welding is scanned in one direction at 4 kHz and in the perpendicular direction at a mixture of 4 and 5 kHz. The rhombic-rastered beam can achieve full penetration with a smooth under-bead. The weld is wider but smoother. Lengeler at CERN obtained similar results with a defocused beam [81]. Since that time cavities are always welded with a controlled defocused beam or a rastered beam to give smooth underbeads.

*Other solutions to thermal breakdown: Guided repair*

If there are one or two gross defects in a cavity due to manufacturing errors, these defects can be located by thermometry or optical inspection and removed by mechanical grinding. By repeated application (e.g., four times) of the guided repair method, the accelerating gradient of a 350-MHz single-cell niobium cavity (RRR = 40) was increased from 5 to 10 MV/m [82]. But it is not so easy to eliminate smaller, more frequently occurring defects. Kneisel made similar improvements to elliptical cavities [41]. In 2008, Iwashita at Tokyo University developed a special optical inspection tool [83] which is widely used for 9-cell ILC and European XFEL cavities. It uses a high-resolution camera and specially designed lighting equipment. Once the defect has been located, imaging systems take pictures of the defect inside the cavity with a resolution better than 10 μm.

Using these methods several 9-cell cavities (some from new cavity vendors) were found to show pit-like defects. The heat-affected zone near the weld shows a tendency to form large (~100 microns) voids, likely originating from high stress regions [84]. These voids grow to 200 microns during electropolishing (EP) and retain their sharp edges [85]. Such defects can be removed by global CBP (described below) or by a local grinding tool [86].

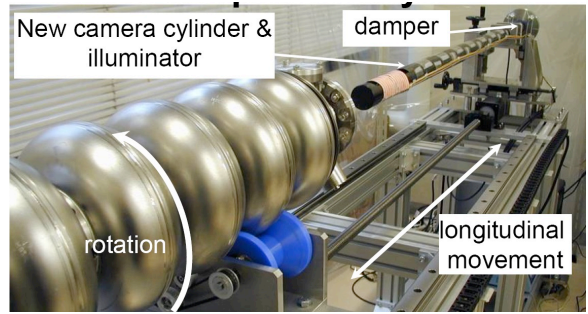

**Figure 9.25:** Overview of the cavity inspection system.

An important step in the preliminary characterization of niobium sheets is eddy current scanning, a non-destructive technique, first developed at DESY by Proch and later improved by Singer [52, 65]. The basic principle is to detect the alteration of the eddy currents with a double coil sensing probe to identify inclusions and defects embedded under the surface. With a sheet size of about 300×300 mm² and a line width of 1 mm, a scan of one sheet lasts about 15 minutes. Pits, inclusions, scratches, and roller marks were the typical defects found, but many of these can be also detected by simple optical inspection. An interesting inclusion found with eddy current scanning in the European XFEL production run was later identified to contain Cu and Fe using Synchrotron Radiation Phosphorescence Analysis [53].

The conventional eddy current system has limited sensitivity (~ 0.1 mm depth). Later, SQUID detectors for measuring the eddy current's secondary magnetic field improved sensitivity and provide excellent signal/noise ratio [53, 87-88].

*Other solutions to thermal breakdown: Barrel polishing*

In 2000, Saito and others at KEK [89-91] developed Centrifugal Barrel Polishing (CBP) to smooth welds and other cavity defects. CBP will be described in more detail in Chapter 14. After CBP the electron beam welding seam becomes completely invisible (Fig. 9.27). CBP poses an increased risk of hydrogen contamination [91] because the continuous mechanical abrasion of the natural protective oxide layer leads to H pick-up from water in the polishing medium. Fig. 9.27 shows how well a rough electron beam weld can be smoothed using CBP.

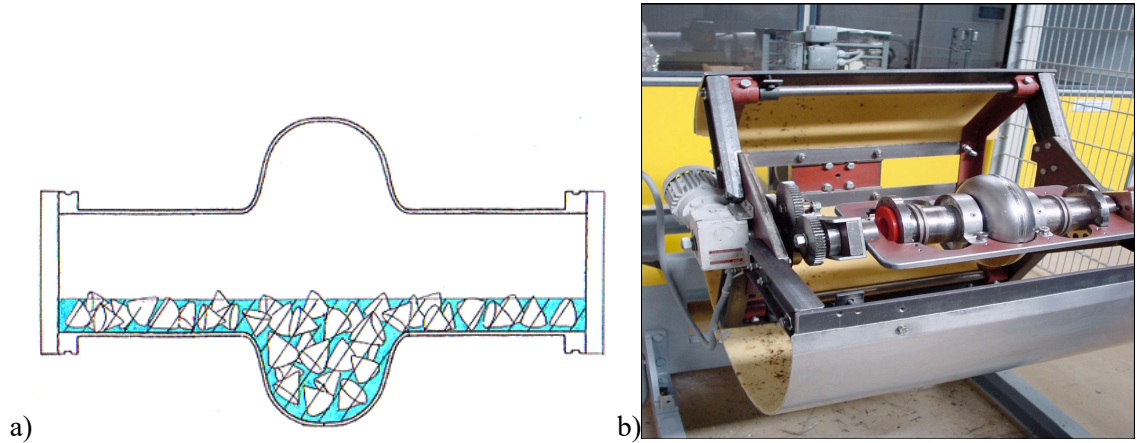

**Figure 9.26:** (a) Centrifugal Barrel Polishing schematic [95]. (b) Single-cell set-up [90].

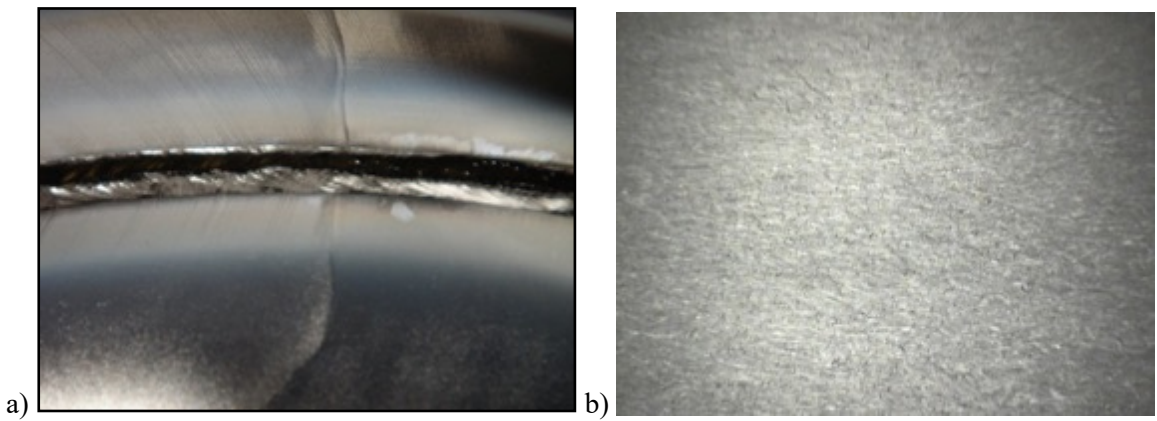

**Figure 9.27:** (a) A rough inside surface weld and (b) its improvement after CBP at KEK.

## 9.7 Thermometry-based diagnostic advances

As pointed out in various sections, thermometry-based diagnostic systems played a key role in improving understanding of the field limitations of multipacting, thermal breakdown and field emission (discussed in the next section). Here we review the progress of diagnostic methods and how the findings advanced understanding.

In 1972, Lyneis et al. detected thermal breakdown locations in single cell cavities [92]. The heat pulses were detected by fifteen 56-Ohm, 1/8-watt carbon resistors. Carbon is a semiconductor and increases its resistance $R$ exponentially with decreasing temperature $T$. At liquid helium temperatures the gradient $dR/dT$ is several tens of Ω/mK, permitting sensitive temperature excursion measurements. The mounting device in the shape of an arm held the resistors a few millimeters away from the cavity wall. The arm of resistors could be rotated azimuthally around the cavity axis so that the location, size and shape of the hot spot from a quench could be measured. In 1977, Lyneis used fixed thermometers in a ring around the S-band cavity to detect heating from multipacting [28], already discussed in Section 9.4.

Padamsee in 1977 [29] confirmed with a 100-Ohm carbon-resistor-based thermometry array (Fig. 9.28) that multipacting induced heating was taking place near the center of the muffin-tin cavity bottom, equivalent to the equator of the cylindrically symmetric HEPL pill-box geometry. Fig. 9.28 shows the location of 138 thermometers in a rectangular frame placed outside a 2-cell cavity. The typical temperature fluctuations during multipacting measured just above the helium λ point are shown in the figure along with the RF reflected power.

Padamsee found it essential to look at temperature signals above the λ point as the sensitivity of the bare thermometers went to zero in superfluid helium. Below the λ point the technique works well only for locating spots during quench when a large heat pulse propagates into the helium bath. Fig. 9.28 shows the location of thermometers, along with heating intensity (as number of bars on the thermometer). When the bath surrounding the cavity is in the normal state of helium the effective thermal conductivity of the helium bath is low. As a result, the temperature excursions on the outside wall due to RF losses are enhanced, and the resistors are more sensitive to small changes in temperature at the outside of the cavity wall. The resistors are also able to detect any bubbles generated at the walls from the heating. The technique was also used to detect heating in lossy spots (weld defects for example) prior to breakdown as well as heating due to the impact of electrons on the walls of the cavity.

The upper part of Fig. 9.28 shows the strength of the heating detected at the bottom of the cups when a significant number of electrons are present inside the cavity ($E_{eff}$ = 4 MeV/m) during multipacting (discussed in Section 9.4). The oscillations observed in the heat pulses (Figure 9.7) are semi-synchronous with oscillations in the stored energy as seen at the bottom of the (reflected power) filling pulse. The largest heat pulses (many dashes) shown in Fig. 9.28 identify the end points of the trajectories of multipacting electrons.

Fig. 9.29 shows typical heat pulses measured by a thermometer outside a weld defect. At low fields (part B) there is a small pre-heating of the defect before the quench. At the quench field value (part A), the pre-heating is stronger, and a large heat pulse is generated during quench.

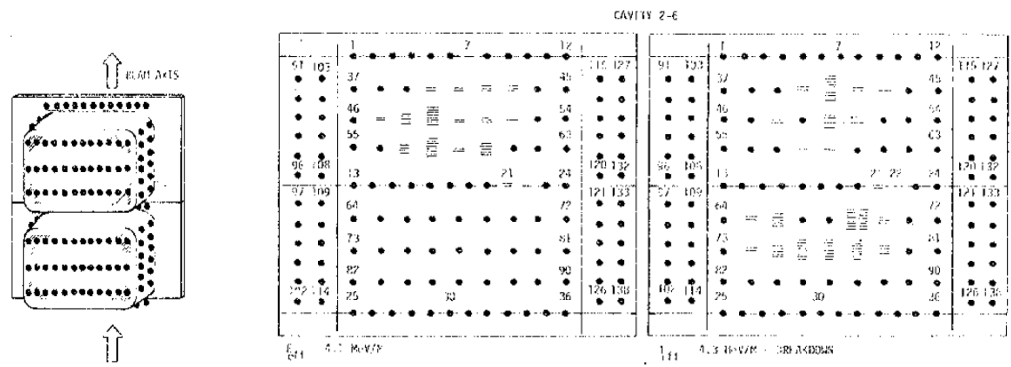

**Figure 9.28:** Left panel: A frame of 138 carbon resistors placed outside a 2-cell cavity. Middle panel: Maps showing placement of resistors (circles) and distribution of heat pulses (dashes) detected outside the cavity during multipacting at 4.1 MV/m. The larger the number of dashes the higher the temperature rise. Resistors # 49 and 59 shows the largest heating (18 mK). Right Panel: Quench at 4.3 MV/m. Resistor #69 shows 54 mK temperature rise.

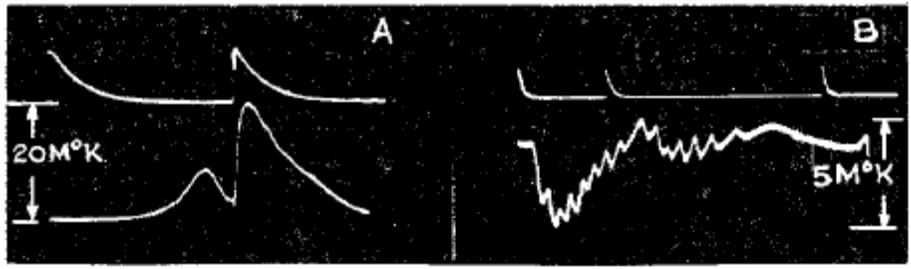

**Figure 9.29** Typical heat pulses observed at carbon resistors placed at a weld defect. Part B shows the preheating of 5 mK at a low field. (Oscillations are probably due to bubbles). Part A shows the large temperature excursion during the quench. The upper trace in both parts A and B show the rf reflected power.

In 1980, Piel [93] initiated a global temperature mapping scheme (Fig. 9.30) for a large 500 MHz single cell cavity. The system, developed at CERN, used a rotating arm of thermometers submerged in subcooled helium (discussed below) that circled the cavity. To minimize the number of thermometers required, a single thermometer bearing arm with 39 thermometers circled the apparatus. The thermometers could glide on the cavity wall under spring tension provided by spring fingers. The carbon resistors (see Fig. 9.30(a)) used were 56 or 100 Ohm (1/8 W or 1/4 W) Allen Bradley resistors, the bakelite insulation of which was ground off to increase sensitivity. A computer system measured the resistance of each thermometer and converted the values to temperatures. The temperature profile along one azimuth of the cavity thereby was obtained in a single scan. The arm was then moved by a motor to a new position.

During a quench all the energy stored in a cavity dissipates at the quench spot so that a substantial heat flux develops leading to a marked increase of the temperature of the helium film close to the quench area. This can be detected easily both in superfluid helium and does not require the resistor to be in contact with the cavity wall. A temperature map of the surface of a cavity below the breakdown field, however, reveals more information about the nature of lossy areas. Such temperature mapping can only be done for bath temperatures above the $\lambda$-point temperature, as discussed above. Piel discovered that the thermometer response was much higher in a subcooled helium bath. Not only was the bath temperature held slightly above the $\lambda$ point but a bath overpressure pressure of 1000 mbar was also applied. In such a subcooled bath, the overpressure suppresses bubbles so that there is no heat transport by bubbles. This reduces the cooling capability of liquid helium and increases the temperature of the outer surface of the cavity near the hot spots. Fig. 9.31 shows one of the first 3-D temperature maps of a superconducting, 500 MHz, niobium cavity operated at an effective accelerating field of 3.2 MV/m. In this early experiment at CERN the clean-room handling was not as well developed as today so that already at an accelerating field of 3.2 MV/m one observes strong field emission heating (discussed in Section 9.8).

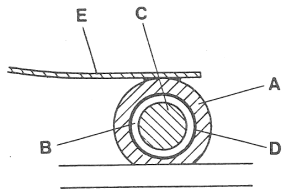 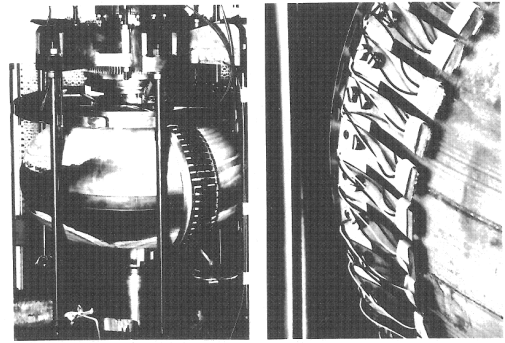

(a)                                    (b)                        (c)

**Figure 9.30:** (a) Cross-section through carbon thermometer for temperature mapping in subcooled helium (b) Thermometer arm bearing 39 resistors making good contact with cavity surface by spring fingers. (c) Close-up view of temperature scanning system used by CERN.

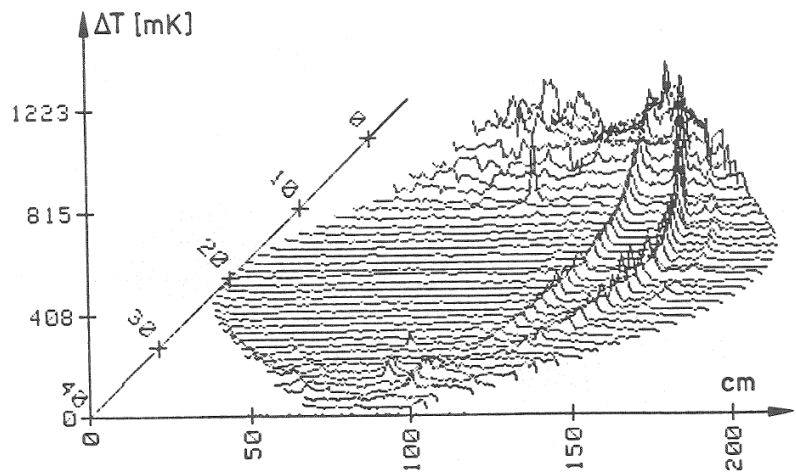

**Figure 9.31:** Temperature map with rotating arm of thermometers in subcooled He for a 500 MHz single cell cavity at 3.2 MV/m. The map clearly shows the line like losses from bombardment of field emitted electrons.

There are a few drawbacks of the rotating arm system operating above the λ point in the subcooled state. One is the long acquisition time (half to one hour) of a complete temperature map for a 360-degree rotation of the arm. Therefore, only stable RF losses in the cavity can be monitored in the steady state. Processing events or thermal breakdown can be observed best if they take place at one azimuth of the rotating arm. Another disadvantage is that the sensitivity of thermometers diminishes significantly when operated below the λ-point of helium. Therefore, the system is not best suited to studying high frequency (> 1 GHz) cavities which when operated above 2.2 K have significant BCS losses that dominate the heating.

These problems were solved in 1987, when Padamsee [94] used an array of 684 thermometers sensitive in superfluid helium and placed at *fixed locations* to cover the entire cavity surface of a 1.5 GHz single cell cavity (Fig. 9.32). Such a fixed system was used in superfluid helium by making improvements in thermometer design (discussed below) and by using Apiezon N grease between the cavity wall and the thermometer. The layer of grease shields the sensing element from the superfluid helium, permitting direct measurement of a fraction of the cavity wall temperature.

A room temperature multiplexing system was used to obtain a fast temperature map, allowing both transient and steady state effects to be studied over the entire surface in one map. Fixed thermometry is inherently much faster than rotating systems. Temperature maps were acquired in 15 s. The main drawbacks of a fixed systems were the many, many wires needed (ribbons and ribbon connectors were used) and the long time and labor required to attach the large number of thermometers to the cavity (about two man-days for a 1.5 GHz single cell cavity).

Substantial improvements in thermometry sensitivity are needed for carrying out measurements in superfluid due to the high heat transfer between the Nb wall and the cooling bath. For example, at heat flux densities of 1 mW/cm$^2$, the theoretical temperature rise at the outer wall drops from 130 mK in subcooled He at 2.2 K to only 7 mK in superfluid at 1.5 K, assuming the Kapitza resistance of unannealed reactor grade Nb for heat flow calculations. One key ingredient for improving thermometer sensitivity was to isolate the carbon element of the thermometer from the superfluid. Carbon resistors were embedded in a G-10 epoxy housing (Fig. 9.33) and sealed with Stycast epoxy, known to be impervious to superfluid He. Thin manganin alloy sensor leads were used to reduce heat transfer to the bath. Another factor essential to high sensitivity was to establish intimate contact between the sensor and the outer surface of the cavity. The surface of the thermometer assembly was ground plane until the carbon element was just exposed. Subsequently the carbon was electrically insulated by several layers of GE-varnish. Apiezon N grease was applied between the thermometer and cavity surface and each thermometer assembly was pressed against the cavity wall with a Be-Cu spring loaded contact pin. Wiring for the large array was made compact by printed circuitry (Fig. 9.32).

To acquire a temperature map, the resistors were scanned once with the RF turned off, and a second time with the RF on. The bath temperature was also measured at the start and stop of each scan to correct for bath drifts. The entire data acquisition process was complete within 15 s. A typical 3-D display showed heating due to field emission at one azimuth as well as defect heating (below quench) as shown in Fig. 9.34 [94]. Maps were taken in rapid succession at increasing field levels, so that field level dependent phenomena were studied in detail.

In a later (1995) upgrade, Knobloch [95] improved the system with better wiring, faster multiplexing and 756 thermometers for a single cell 1.5 GHz cavity. The overall sensitivity and accuracy were improved (down to several μK) and the acquisition time reduced to a few seconds per map. Knobloch made many important discoveries about defects, two-surface multipacting, field emission and residual resistance with the high sensitivity system [51]. The improved system has been duplicated at DESY by Pekeler (see for example, temperature maps in Section 9.9 on HFQS), JLab by Ciovati [96] and Fermilab by Romanenko [97] to study 1.3 and 1.5 GHz cavity phenomena with interesting new discoveries from all these applications.

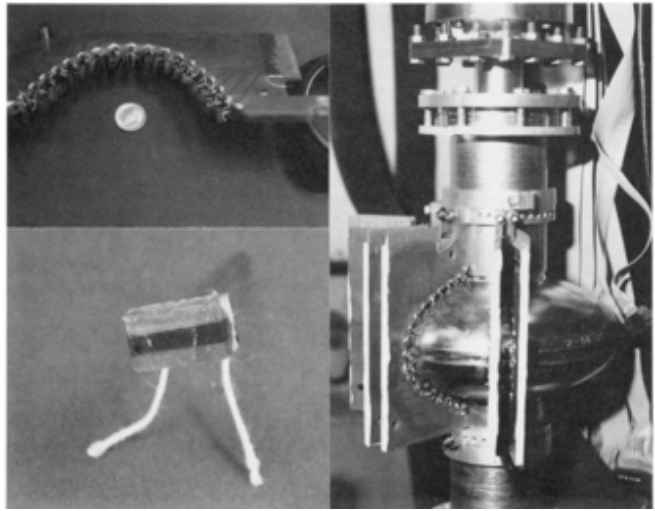

**Figure 9.32:** Single cell 1500 MHz cavity with printed circuit boards bearing 19 thermometers sensitive in superfluid helium. An individual thermometer assembly is shown on the left. Boards are placed at 10° azimuthal intervals.

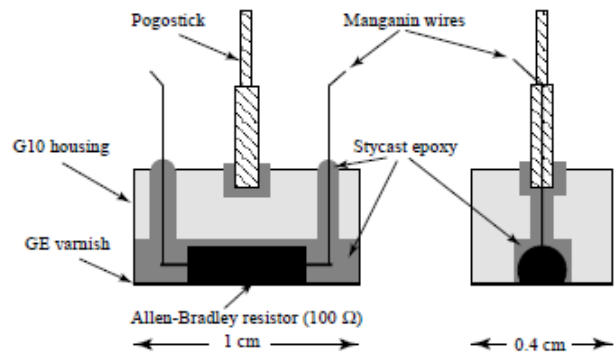

**Figure 9.33:** Individual thermometer assembly construction to achieve sensitivity in superfluid helium.

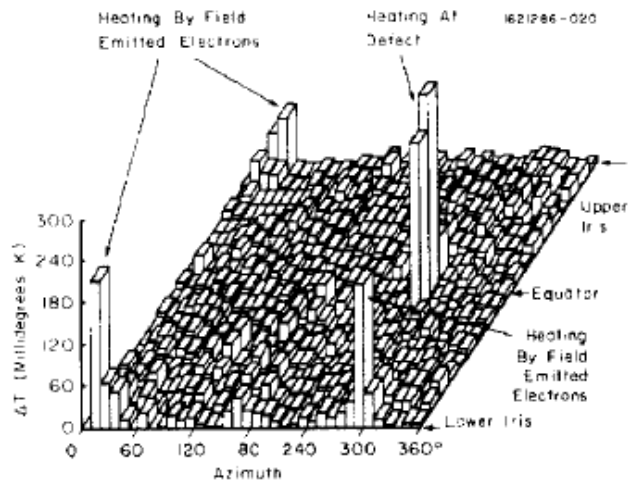

**Figure 9.34:** A typical high-speed superfluid helium temperature map showing heating by field emitted electrons at two to three locations near the iris, and at a defect on the equator. Surface $E_{peak}$ = 15 MV/m.

*Quench detection with second sound in superfluid He*

In the early 1970s researchers at HEPL originated a novel technique based on second sound propagation in superfluid helium [14]. They used an array of 14 resistance thermometers distributed along the length of the 7-cell structure to measure the time of arrival of the heat pulse initiated by breakdown. The point of origin could be established within ±1 cell of the structure. In 1979, Shepard [98] at Argonne National Laboratory used second sound to locate quench spots in low-beta resonators (Fig. 9.35). He used several germanium resistance thermometers housed inside a capsule and placed these inside the niobium tube of the split-ring resonators. He was able to narrow down the quench location. Fig. 9.35 shows the second sound signal.

In 2008, Conway and Hartill [100] further developed the method at Cornell by measuring the time of arrival of the second sound wave at three or more detectors (Fig. 9.36), to locate a quench-causing defect with a spatial resolution of about (~1 cm). The detectors were oscillating superleak transducers (OST) which measure the fluctuating superfluid helium counter flow velocity to detect the time of arrival of second sound waves [99-102]. The OST elements were parallel-plate capacitors with one rigid plate and one flexible-porous plate. The pore diameter was chosen to clamp the flow of the normal fluid while allowing the superfluid, with zero viscosity, to pass freely. The arrival of a second sound wave at the OST causes the flexible-porous plate to move with the normal fluid as the second sound wave passes. The capacitance of the detector is continuously monitored to measure the arrival of the second sound wave. A typical OST arrangement uses 8 transducers evenly distributed around a cavity. This became a cost-effective and simple method to determine quench locations.

The method was used to locate a pit defect on the equator weld of a reentrant 9-cell cavity, reported in [100]. This cavity was subsequently tumbled, removing just enough material to eliminate the weld pit. After reprocessing, the cavity accelerating gradient improved from 15 MV/m to 30 MV/m.

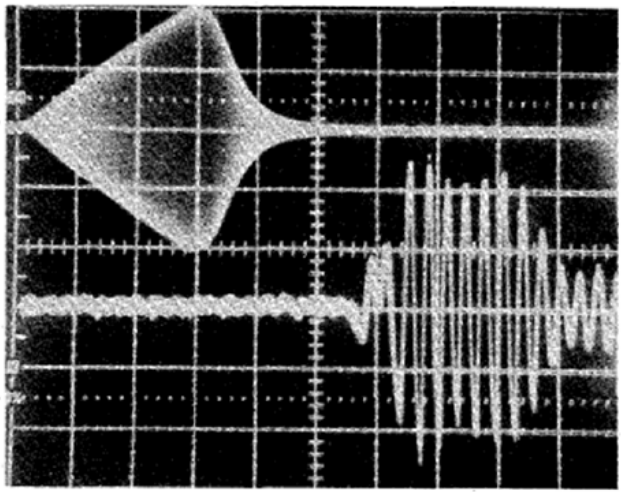

**Figure 9.35:** Shows an oscilloscope photograph of the second sound quench detection method [98]. The top trace is the RF field level in split-ring resonator, which was excited to $E_{acc}$ = 3.0 MV/m where the cavity quenched. The lower trace is proportional to the temperature of a thermometer located inside the split-ring resonator.

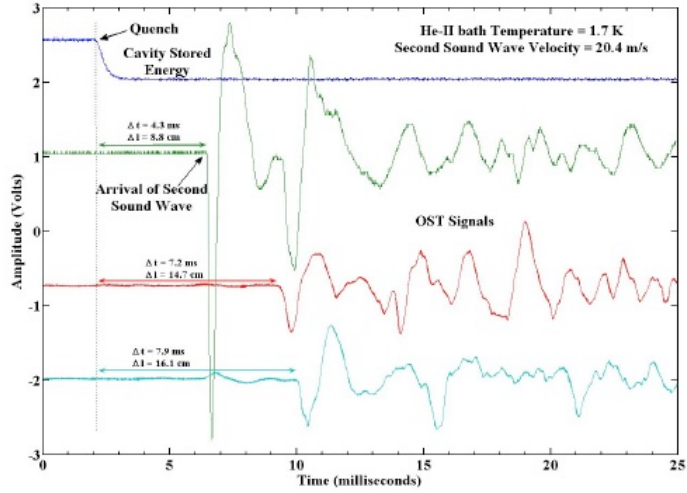

**Figure 9.36:** A typical quench event observed with three different transducers. The top trace is the amplitude of the cavity RF field. The lower three traces are the second sound signals measured with 3 distinct OSTs. The variation in the second sound wave time-of-arrival is well correlated with the variation in propagation distance to each transducer [99].

## *9.8 Field emission*

### *Understanding field emission*

After thermal breakdown came under control in the 1990s, field emission (FE) [103, 104] took over as the dominant limitation at fields above 10 – 15 MV/m. Signs of field emission in the best performing cavities were already familiar since 1970s, as strong *Q* drops and intense X-ray emission. Field emitted electron currents were also found to be limiting phenomena in high voltage DC devices. The exponential nature of field emitted current is characteristic of the quantum mechanical tunneling process first calculated by Fowler and Nordheim (FN) [105].

$$I_{FN} = 3.85 \times 10^{-7} A/V^2 A_{FN} \frac{(\beta_{FN} E_{em})^2}{t^2(y)} \times \exp\left(-5.464 \times 10^{10} V/m \frac{v(y)}{\beta_{FN} E_{em}}\right)$$

Here *t(y)* and *v(y)* are tabulated functions that account for the image charge effect, and are approximately unity. The observed emission current follows the modified FN law, where the electric field at the emission site, $E_{em}$, is replaced by a locally enhanced field, $\beta_{FN} E_{em}$, where $\beta_{FN}$ is the FN field enhancement factor. Since the FN theory calculates the current density, an effective emitter area, $A_{FN}$, is also required to determine the field emission current, $I_{FN}$. Both $\beta_{FN}$ and $A_{FN}$ are empirical parameters, helpful for characterizing the intensity of the field emission, but their physical meaning is debatable. Typical $\beta_{FN}$ values are between 50 and 700 (for very strong emission), and the emissive area values are between $10^{-9}$ and $10^{-18}$ m² [106, 107]. The large spread in the FN parameters is probably due to many physical factors that determine the nature of the emitter, some of which are understood and discussed, and some are still under investigation.

In 1974, Turneaure, BenZvi [26] and others studied field emission from 1.3 GHz cavities using NaI(Tl) crystal scintillator to measure the X-radiation produced by field-emitted electrons and make X-ray photographs (Fig. 9.5). Advances in temperature mapping at CERN in 1981 (see Section 9.7) showed that emission generally arises from particular spots, called "emitters," located in high electric field regions. The emerging electrons travel in the RF fields of the cavity, impact the surface and deposit most of the energy gained from the RF fields as heat, and some as X-rays. Much was learned about the nature of emitters from such cavity studies in which the emitters were located by temperature maps and their field enhancement factors were characterized from temperature increments and X-ray emission.

Under encouragement from CERN by Piel and Lengeler, Fisher and Niedermann [106] at the University of Geneva built a special apparatus (Fig. 9.37) to characterize the nature of DC field emission in Nb. They devoted considerable resources to understanding the origins and nature of field emitters. An important feature discovered was that the density of field emission sites increases (generally exponentially) with field [106] (Fig. 9.38). Thus, not only does the field emission current grow exponentially with field, but also the number of active emitters.

Once an emission site was identified in the DC field emission scanning apparatus, it was studied by the surface analytical tools incorporated in the field emission scanning device, or by on-line instruments. Fig. 9.39 shows an electron microscope (SEM) picture of a carbon flake emitter found in the Geneva DC studies. A particle was almost always found at an emission site. A large number of emitters were subsequently studied with the SEM. Micron and sub-micron size contaminant particles, usually "metallic" (conducting or semiconducting) were identified as the dominant sources of field emitters. Emission properties of many hundreds of emitters were characterized. When a region free of particles was probed, there was no intrinsic field emission up to 200 MV/m, which was encouraging for future efforts to reach accelerating fields up to 100 MV/m.

One of the most important results was that the speculated "sharp whiskers" expected to be responsible for strong emission and corresponding high $\beta_{FN}$ values were not found at emission sites. There was also no correlation found between the location of emitters and grain boundaries, ruling out the frequently quoted possibility that a step at a grain boundary is a cause for field emission via geometric enhancement. Certainly, if the surface is scratched, the sharp projections at the edge of the scratch are field emitters [108]. However, scratches are rare on carefully prepared surfaces.

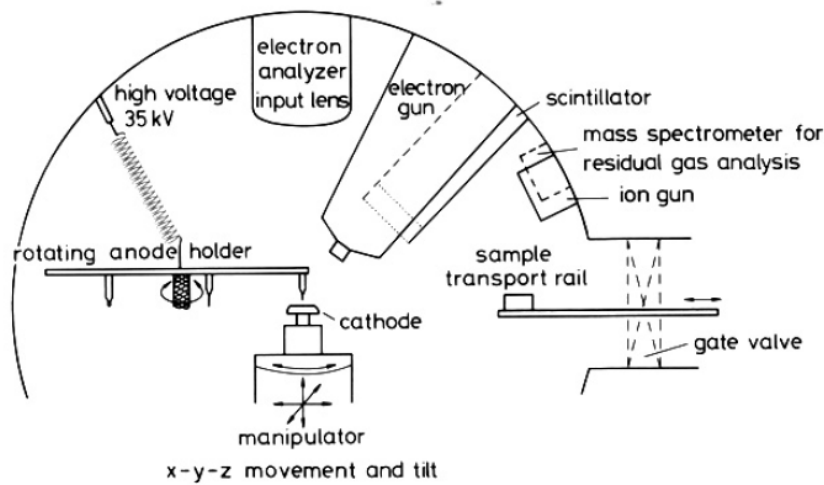

**Figure 9.37:** Apparatus for scanning a surface with a high-voltage needle and analyzing emission sites.

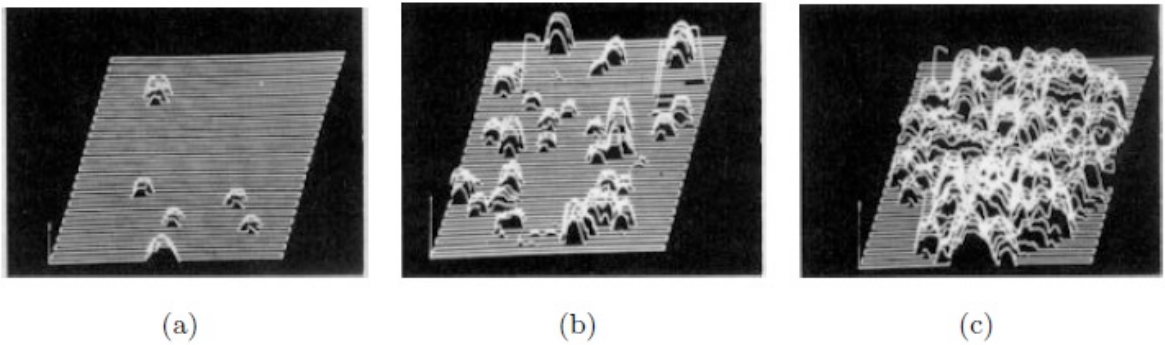

**Figure 9.38:** Emission sites found on a cm² sample of niobium using DC fields (left) 50 MV/m, (middle) 90 MV/m, (right) 100 MV/m. Note the large increase in number of FE sites with increasing field.

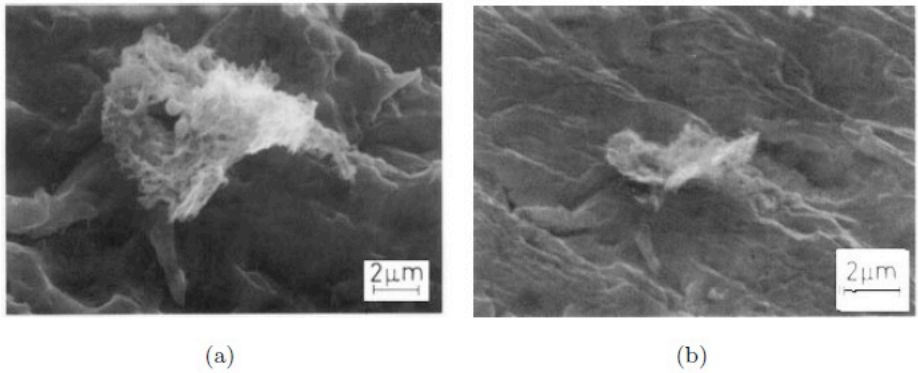

**Figure 9.39:** (a) An emission site analyzed to be a carbon particle (b) same as (a) but viewed at a different angle [106]. Note the sharp or jagged features of the particle.

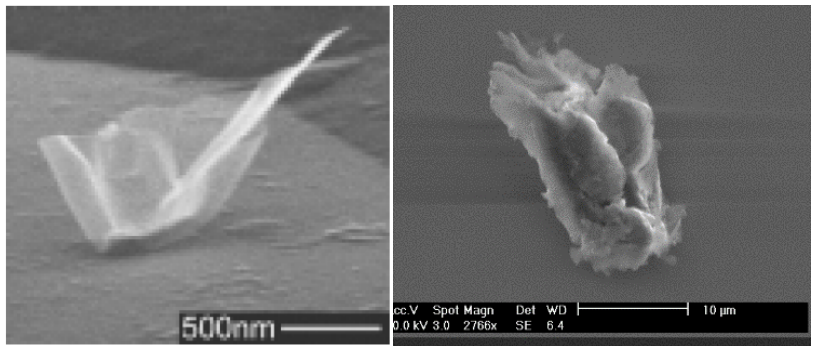

**Figure 9.40:** Field emitting particles. (a) Sub-micron field emitting particle found on sample prepared with 9-cell cavity [115] (b) Al particle found at a field emission site in the dc field emission scanning apparatus and subsequently analyzed with the SEM [116]. Note the jagged and sharp features of the particles.

Fig. 9.37 shows the pioneering Geneva field emission scanning apparatus using DC fields up to 200 MV/m in a UHV environment. A similar apparatus was used at Wuppertal University [107].

Artificial emission sites were introduced to improve understanding of the nature of field emitters. These were most often carbon, iron, nickel, molybdenum disulfide, alumina, and silica particles. Such DC field emission studies were continued by Bonin at Saclay [108] and by Mahner [107] at Wuppertal. Observed emitters almost always turned out to be "metallic" microparticle contaminants, i.e., usually electrically conducting.

Bonin found that not all foreign particles present on a surface are necessarily active emitters. Studies showed that less than 10% of particles present are emitters up to maximum fields ~100 MV/m. Emitting particles tend to have irregular shapes with jagged features that are likely responsible for some of the local field enhancement. Artificially introduced nickel particles showed strong FE if the particles were jagged, but no FE when the particles were smooth (Fig. 9.41).

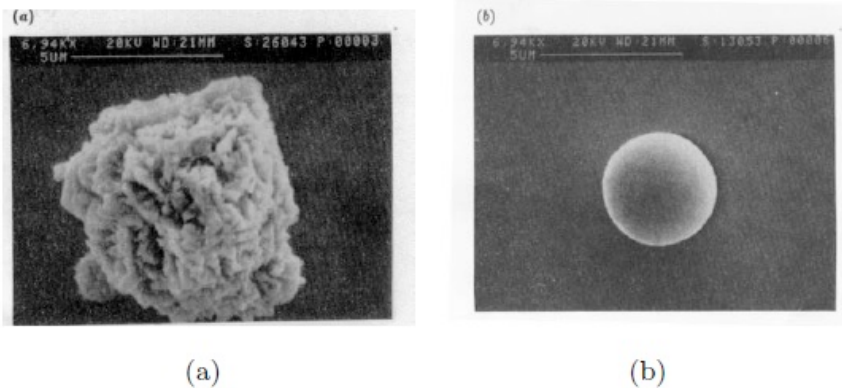

**Figure 9.41:** SEM micrographs of two nickel particles from an artificial field emitter study: (a) jagged shape and emitting (b) smooth and non-emitting up to 100 MV/m.

A simple interpretation for a large FN enhancement factor would be that the particle as a whole enhances the local field by a factor of about 10, and smaller protrusions on the particle further enhance the field by another factor of about 10 [108]. The product would be sufficient to explain observed beta values of 100. But higher values of $\beta_{FN}$ demanded other explanations. Static electric field calculations for a tip-on-tip geometry support the idea that field enhancement factors can be cascaded. All the emitting particle pictures presented do indeed show jagged features. Microtip features also offer a simple explanation of FE current instabilities; namely, when one tip melts and becomes smooth, the local beta FN value decreases, and emission from another tip takes over.

Turning to emitter studies in RF cavities, at Cornell, Graber and Padamsee [109] conducted RF tests on 3 GHz single cell cavities with thermometry to locate emitters, followed by a dissection of the cavity to examine emitters in the SEM followed by EDX analysis for elemental identification. Knobloch and Padamsee [110] continued similar studies on 1.5 GHz cavities. In Fig. 9.42 a temperature map taken [111] at $E_{pk}$ = 17.2 MV/m shows several active emission sites near the iris of the cavity (locations 4, 5, 6 and 7). After the temperature map identifies the interesting sites, the cavity is dissected in a class 1000 clean room, and the surface is examined in an SEM. During dissection, which is carried out with a large, clean pipe-cutter, the cavity is pressurized with filtered nitrogen to prevent dust contamination. Vacuum suction is maintained near the cutting tool. Fig. 9.43 shows particles containing Fe and Cr (probably stainless steel) at the predicted emission site (#4). Note that there are a couple of small balls present in the outlined region, suggesting that the particles suffered local melting. We will discuss partial melting of emission sites later. The rest of the site is a collection of jagged particles. Following the initial SEM examination, the area was cleaned with a high-pressure carbon dioxide jet to ensure that the particles were not loose debris that landed on the cavity during dissection, despite the precautions taken to avoid dust. Reexamination in the SEM showed that the site was unaltered, except that the molten balls were missing. It is clear that all the stainless particles had strongly adhered to the cavity surface. The balls became weakly attached to the surface due to necking upon solidification, so that the balls were dislodged by the gas jet.

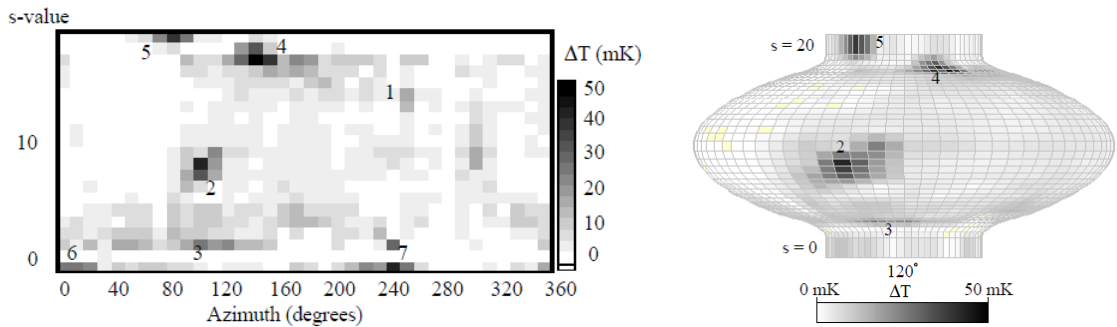

**Figure 9.42:** Temperature map of a 1-cell, 1.5-GHz cavity showing several hot spots, some of which are field emission sites. At $E_{pk}$ = 17.2 MV/m the cavity shows several emission sites near the iris of the cavity (locations 4, 5, 6 and 7). After cavity dissection, the emitter at location #4 was analyzed in the SEM.

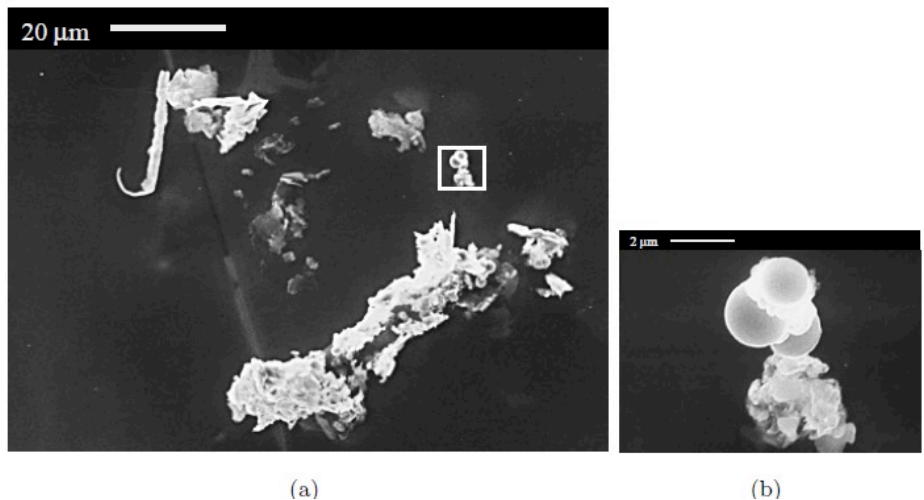

**Figure 9.43:** (a) SEM micrograph of particles at site #4 of the previous figure. (a) The field emitting particle was found with thermometry followed by dissection of a 1.5 GHz cavity. Carbon, oxygen, iron, chromium, and nickel were among the foreign elements detected. Note the sharp features on the particles. Note also the cluster of small spherical balls in the framed portion which indicate that a part of the site melted. (b) The melted cluster is expanded. EDX analysis show that the particles are stainless steel.

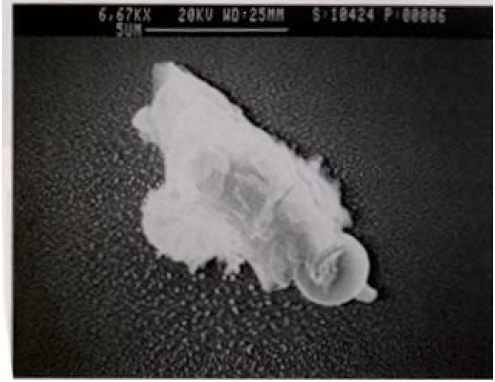

**Figure 9.44:** SEM micrograph of an indium metal flake field emitter. A small melted region can be recognized by its spherical shape. The particle was subjected to a maximum electric field of 26 MV/m in the RF test of the 3-GHz cavity in which it was found.

Fig. 9.44 shows another emitter [112] caught in field emitting action, this time an indium particle which was found in a one-cell, 3-GHz cavity at the location of a field emission site. This particle does have jagged features. Like the stainless particle, the indium particle also shows a small molten spot. A possible reason for these melted regions on the emitters is RF heating, but this would melt the entire flake. The more likely explanation is that the stainless steel and indium particles field emit only from a small region, which heats up due to Joule heating from the emission current. At a high surface field, the exponentially increasing emission current can melt the superficial particle. It is possible that when one region melts, it becomes smooth, so emission from another region takes over. This is one possible explanation for the observed instability in the emission current, when the fields are raised for the first time. Melting also implies the origins of a gas cloud formation near the emission site, which could later lead to an RF spark and processing.

In cavities, many of the particulate contaminants can be traced to various stages of sample preparation and assembly. For example, cavity assembly tools, vacuum pipes, and weld joints of vacuum pipes are possible sources of emitters with iron and stainless. Carbon and silicon are likely due to airborne dust particles. Taking these lessons to heart, cavity assembly everywhere has moved to Class 10 – 100 clean rooms. Increased vigilance in particulate cleanliness during final surface preparation and assembly procedures is essential to keep foreign particulate contamination and associated emission under control.

### *Condensed gas enhancement of FE*

There are several experiments [113] which show that condensed gases activate field emission, presumably by adsorbing on the surface of dormant particulate sites. The emission landscape observed by temperature maps was occasionally found to change on warming a cavity to room temperature and re-cooling. Two emission sites apparent in Fig. 9.45 (a) are no longer active, or become dormant, after cycling to room temperature, and cooling down again to 1.6 K, as shown in Fig. 9.45(b). However, when helium gas was admitted into the cold cavity, one of the two former sites reactivated as shown in Fig. 9.45(c).

In another experiment, an emission-free superconducting cavity (Fig. 9.46(a)) was exposed to a steady stream of oxygen gas while the cavity was cold. Most of the oxygen probably condensed on the vacuum pipes leading to the cavity, but some gas reached the cavity surface, as evidenced by a sudden increase in field emission, accompanied by a drop in the $Q$ and the field. Fig. 9.46 (b) shows a strong emitter activated by the condensed gas. To rule out the possibility that the new emitter was a new particle introduced accidentally with the oxygen stream, the cavity was cycled to room temperature. On returning to 2 K, emission at the previously activated site was absent. However, readmission of oxygen reactivates emission at the *same* site. It is highly unlikely that a particle would land on the same spot as two separate doses of gas are introduced.

At CERN, cavities were exposed to a mixture of gases (H, $H_2O$, CO, and $CO_2$) typically found in an accelerator vacuum system. The onset of field emission changed from 10 MV/m to 7 MV/m if more than one monolayer of gas was adsorbed [114].

*Cures of field emission*

By and large, both RF and DC studies discussed above revealed that emitters are micron- to sub-micron size contaminant particles. The studies promoted increased vigilance in cleanliness during final surface preparation and assembly. Class 100 (or better) clean room environment is important to keep particulate contamination and associated emission under control. New approaches were adopted to strive for a higher level of cleanliness in cavity surface preparation, leading to fewer emission sites and better cavity performance. But clean room assembly alone did not have the drastic impact needed for FE reduction. Some additional attack was needed.

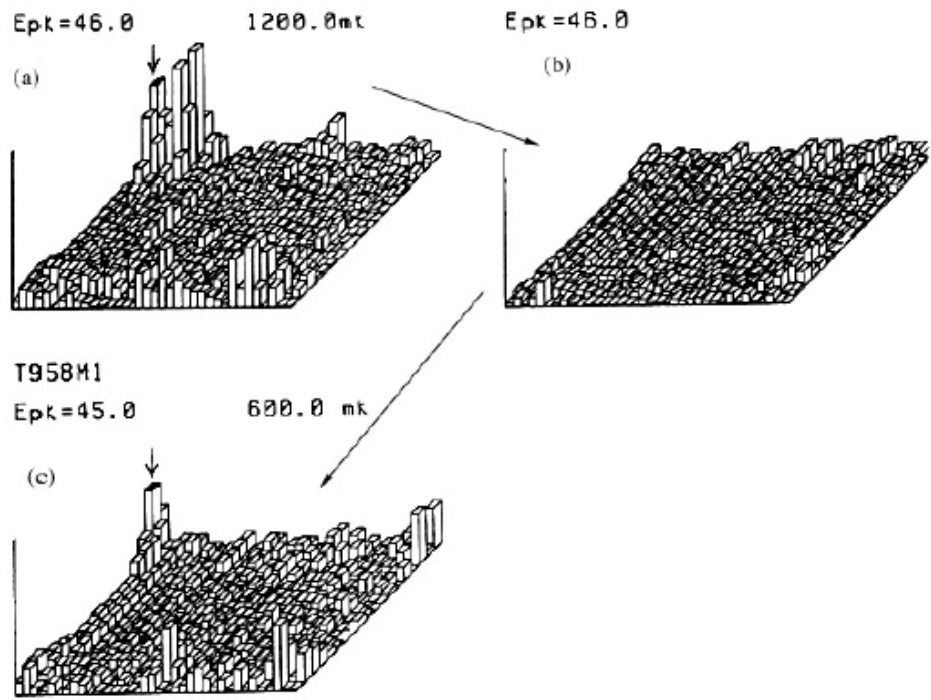

**Figure 9.45:** Temperature maps from a 1.5-GHz single-cell cavity. (a) Several field emitters are active at $Epk$ = 46 MV/m. (b) On cycling to room temperature and cooling down, emission disappears, presumably due to removal of condensed gas. (c) On admitting He gas into the cold cavity, field emission reappears at some of the same sites as in (a), presumably due to condensation of gases at the dormant sites [113].

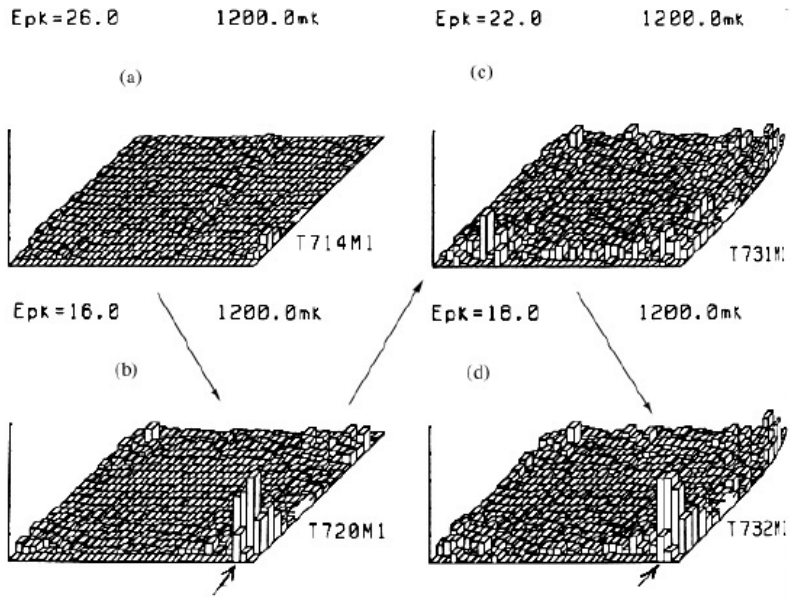

**Figure 9.46:** Temperature maps of a 1500-MHz cavity. (a) The surface is emission-free at 26 MV/m. (b) Emitters appear when the cavity is exposed to oxygen. (c) Emitters are no longer active after cycling to room temperature. (d) The same emitters reappear when exposed to oxygen a second time.

In 1994, Saito at KEK [115] studied high-pressure (100 bar) water rinsing as a promising candidate for removal of particles stuck to the RF surface. The technique was originally tried by Bloess at CERN [116] in 1992, but the potential for field emission reduction in SRF cavities was not clearly demonstrated. Saito studied the effect of high-pressure rinsing using a silicon wafer as the test surface, and laser scanning to detect the superficial particles. Fig. 9.47(a) shows a large number of foreign particles found when a silicon disk was exposed to cavity treatment chemicals outside the clean room, rinsed with water used to clean niobium cavities, and dried in a Class 100 clean room. The idea was to expose the disk to the same environment as the cavities. There are still more than 10,000 particles accumulated over an area of 100 cm$^2$. As the distribution of Fig. 9.47 shows, most of the particles are between 0.3 to 1 μm in size. When a similarly prepared disk was subjected to high-pressure rinsing (HPR) as the final treatment, the number of particles was drastically reduced, as shown in Fig. 9.47(b). A 100-bar jet of ultrapure water dislodges surface contaminants normally resistant to conventional rinsing procedures.

In 1994, Kneisel at JLab [117] demonstrated the full potential of the HPR method for cleaning cavities, and reached high gradients. The benefits of HPR in reducing field emission was well demonstrated in tests on single-cell and 5-cell cavities at TJNAF. In 10 tests on single-cell 1.5-GHz cavities, they reached $E_{acc}$ = 25 MV/m, with 32 MV/m as the best cases. Subsequently, the RRR of several 5-cell cavities at TJNAF was improved to RRR = 500 by titanium solid state gettering to also avoid quench. When HPR was applied to these high RRR cavities, it was possible to overcome both the field emission and quench limitation, to give the excellent results shown in Fig. 9.48 [118]. The maximum field was limited by RF power available.

Similarly, good results came from the DESY 9-cell, 1.3 GHz structures for the Tesla Test Facility (TTF) [119], as shown in Fig. 9.49. In a spectacular best result achieved with HPR, the $Q_0$ remained near $4\times10^{10}$ from low fields all the way up to $E_{acc} = 25$ MV/m. Most cavities appear field emission free up to $E_{acc} = 15$ MV/m.

HPR has also proven effective to re-clean cavities that were contaminated during assembly and continued to show strong field emission during RF tests.

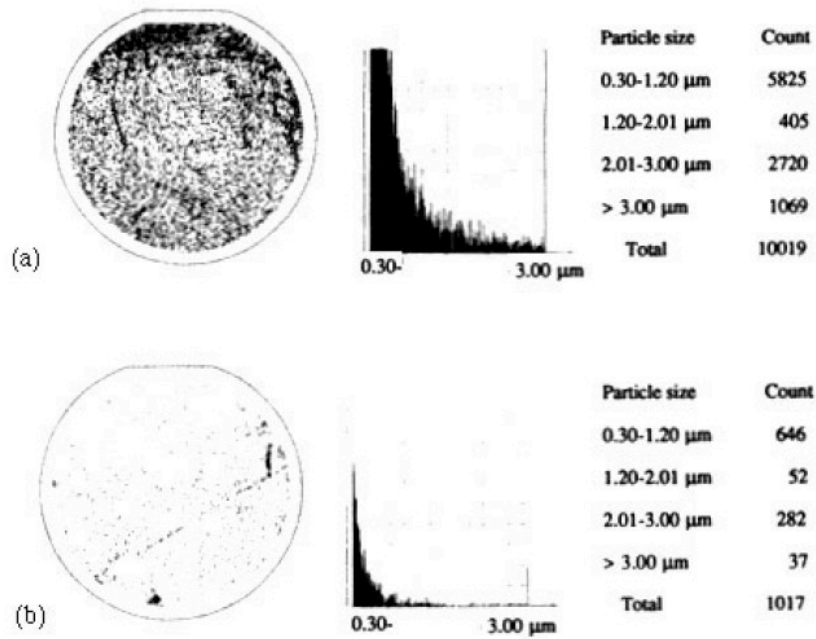

**Figure 9.47:** (a) A 100-cm$^2$ silicon wafer disk prepared by exposure to standard chemicals and cleaning techniques used for superconducting cavities shows a large number of contaminant particles detected by a laser scanner. (b) The same disk, after high-pressure rinsing, shows a substantial reduction in particle count.

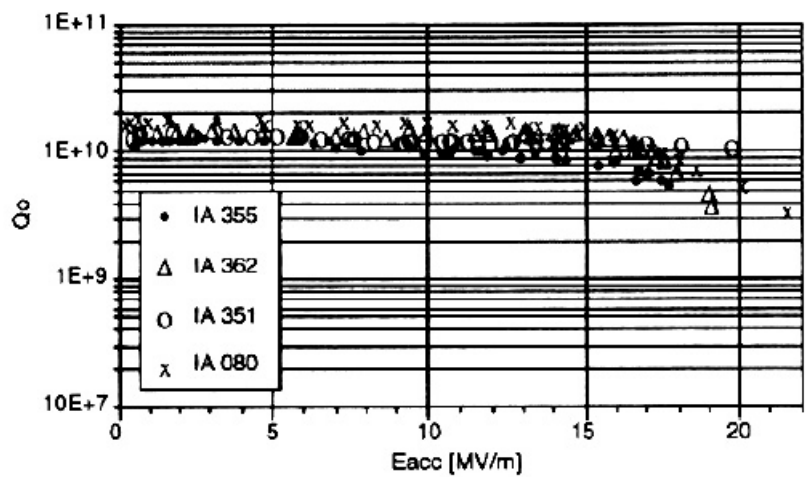

**Figure 9.48:** Field emission-free and quench-free performance of four 5-cell JLab cavities after improving RRR to 500 by solid state gettering with titanium to avoid quench, and final cleaning by HPR to quell FE. The maximum field in these tests was limited by the available RF power.

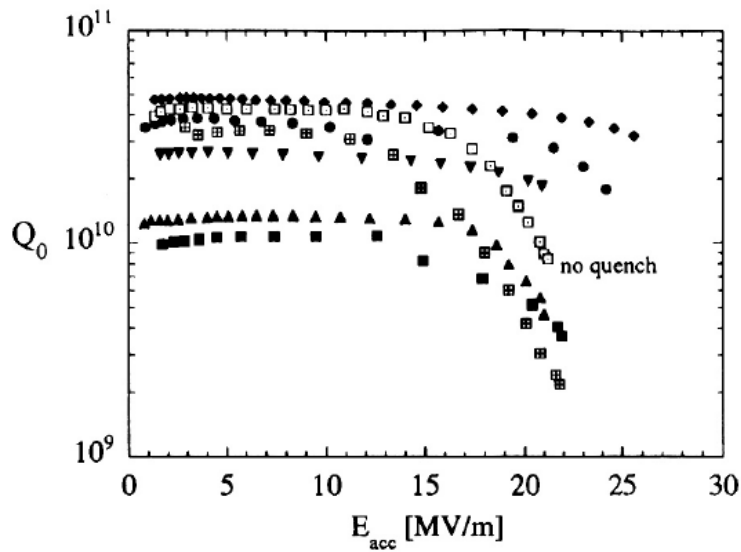

**Figure 1.9.49:** Performance of seven 9-cell TTF cavities after improving RRR to 500 by solid state gettering with titanium and followed by HPR. The maximum field in these tests was limited by thermal breakdown, except for one cavity (labeled "no quench") which was limited by the available RF power.

*Alternative methods for curing field emission*

Before HPR became the method of choice, there was some progress in efforts to reduce field emission, starting with gains established by early techniques, such as helium processing and heat treatment (HT).

*Helium processing and Plasma processing*

Field emission can be reduced by *helium processing,* after introducing a low density of helium gas in the cavity [120]. Many laboratories have tried it with some success. The first experience with He processing was at HEPL in the 1970s under Schwettman. Typically, the cavity is filled with about 1 mtorr of helium, as measured at room temperature. A higher pressure risks RF breakdown of the gas. A systematic study of helium processing for about 300 CEBAF upgrade cavities gave a net gain of about 1 MV/m in onset field for field emission, with best improvements of 3 – 4 MV/m for a few cavities [121].

There are several likely mechanisms for helium processing to reduce field emission. In cases where the field emission decreases within a few minutes after introduction of the gas, processing likely takes place by removal of condensed gas which enhances emission at potential emitting particles [122], as discussed above. In cases where field emission decreases after many hours, Weingarten suggested that processing could take place by removal of emitter material via sputtering [46]. Helium gas can also help trigger a spark discharge, and RF processing of the emitter [123, 124].

Plasma processing at room temperature has been developed by SNS to show an average increase of 2.5 MV/m [125]. As with helium processing, plasma is likely to be effective by removing adsorbed gas layers from dominant emitters. By comparison, the high pulsed power technique

(discussed below) completely destroys particulate emitters in a micro-discharge to give larger benefits.

*RF Conditioning*

Despite the enormous success in controlling field emission, it can occasionally return and limit the gradient after the final assembly of structures, or upon installation of power-coupling or other devices into a cavity string. During these assembly procedures there is generally much human activity so that there is a strong probability of "emitter-dust" falling into the cavity.

Field emission from accidental contamination can be reduced and gradients increased by *"RF processing"*. By raising the surface electric field at the emitter, the emission current increases exponentially, parts of the emitter can melt, and the emitter can be completely destroyed. During RF processing, the cavity may suffer from occasional thermal quenches due to intense bombardment from the high field emission current. RF processing of emission is routinely experienced in all laboratories when raising RF power for the first time while testing or commissioning cavities.

Major progress in the understanding the physics of RF processing came through studies of field emission and processing from many RF cavity and DC experiments at Cornell by Graber, Knobloch, Crawford, Moffat, Werner and Padamsee [51, 103, 104, 109, 110, 124, 126-133]. Characterizing processed emitters using SEM, EDX, and Auger showed that emitter processing with CW RF power is most often an explosive event that accompanies a local "discharge," or "electrical breakdown" of the insulating vacuum near the emission site. The microparticle responsible for emission is destroyed in the discharge, leaving behind a micron to 10-micron size molten Nb crater or craters, surrounded by a 100 – 200 µm "starburst-shape" feature created by the plasma cleaning of the discharge. Monolayer traces of the original particulate emitter are found inside the crater by sensitive Auger methods. The rim of the melted crater after processing is smooth so that it likely does not become another emitter. Similar features (craters surrounded by starbursts) are found in DC voltage breakdown studies where emitters are also processed. Models for the processing event from emission to voltage breakdown have been proposed and numerical simulations carried out [51, 124]. These suggest that the once a critical field emission current is reached the destruction process is very fast (ns).

Graber and Padamsee first analyzed such a RF processing event during CW RF test [109]. When field emission starts, the $Q$ falls sharply due to the exponentially increasing field emission current. Temperature maps record the original field emission, as shown in the temperature map of Fig. 9.50. At about $E_{pk}$ = 29 MV/m, when the incident RF power is increased, the peak field in the cavity jumps to 39 MV/m. There is a familiar processing event. The figure shows the temperature map taken after the processing event. Note that the "before" and "after" maps, which were both recorded at the same field, show that the field emission heating is substantially reduced at 29 MV/m. Upon dissecting the cavity and examining the predicted emitter location in the SEM, the site shown in Fig. 9.51 is found. The 200 µm site has a "starburst" shape with a 10-µm molten crater-like core region accompanied by micron-size molten particles within and near the crater. EDX analysis shows that the starburst region and the molten crater are all pure niobium, within detection limits. The particulate matter in the crater region, visible more clearly in the expanded Fig. 9.51, reveals copper as the only contaminant. Presumably, a µm-size copper particle was originally responsible for the field emission.

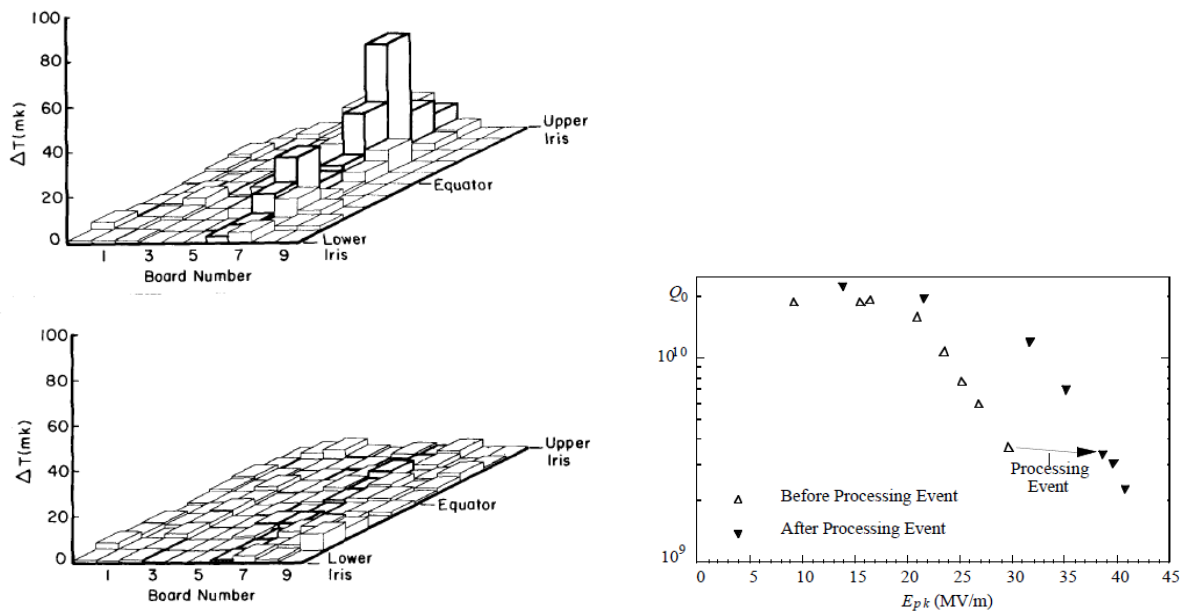

**Figure 9.50:** (Left) Temperature maps of a processing event in a single cell 3-GHz cavity, realized by increasing the CW RF power. (Left-upper) Temperature maps at $E_{pk}$ = 29 MV/m before the processing event show the heating due to field emission. (Left-lower) After the processing event, heating due to field emission is gone. (Right-lower) At 29 MV/m, the $Q$ suddenly increased from $3\times10^9$ to $> 10^{10}$. After the event, it was possible to raise the surface field to 40 MV/m.

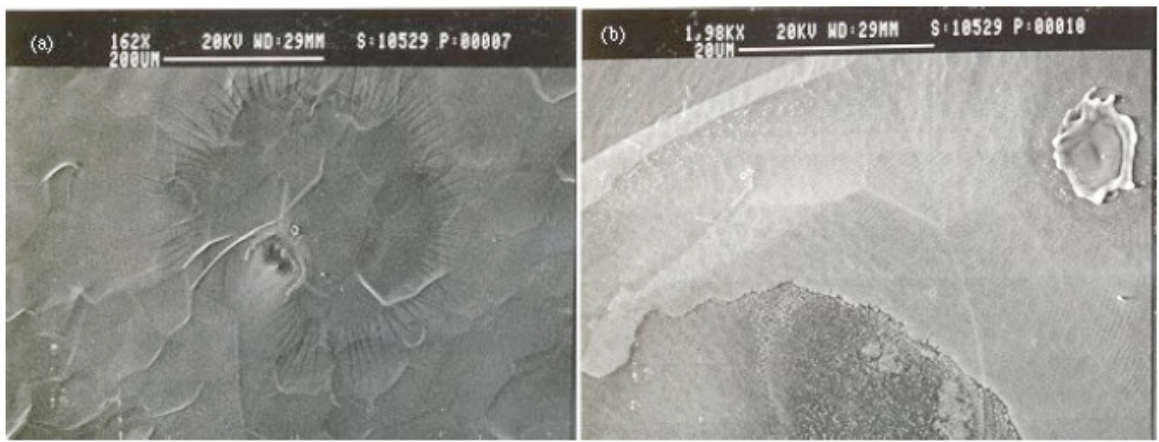

**Figure 9.51:** SEM pictures of the processed site found at the location predicted via temperature maps above. (a) Low magnification and (b) high magnification of the crater region within the starburst shape feature of (a). A particle at the center of the molten crater as well one outside the crater were found to contain copper, leading to the conclusion that the emitter was a copper particle destroyed by RF processing.

The molten crater and splash-type features at the edges of the crater make it clear that emitter processing is an explosive event. The fact that the event can melt niobium on the cold surface suggests that the explosion takes place on a time scale much shorter than the thermal relaxation time. Sensitive Auger studies of the starburst region show that the larger (400 μm) starburst region is much cleaner at the surface, as for example devoid of fluorine, normally found everywhere else on the surface due to the HF acid used for cavity chemistry. Most likely, the plasma formed during the arc was about 400 μm in diameter and the cleaning took place by ion bombardment from the plasma.

Note that field emission decreased, and higher field levels were reached (29 MV/m to 40 MV/m), despite the appearance of molten droplets of Nb in the craters. *The discharge event and craters it causes do not increase field emission,* as may be feared. This is not surprising in view of the fact that studies at Saclay show that smooth particles do not emit (see Fig. 9.41). In some cases, it has been found that processing events can cause a small increase in residual RF losses, possibly due to the spread of the emitter debris. In most cases, however, $Q$ values $\sim 10^{10}$ can be maintained. This example of CW RF processing (used routinely during cavity testing) clearly provides evidence for the nature of RF processing as an explosive event which destroys the emitter. The discharge probably starts with melting of parts of the emitter and accompanying gas generation.

In another possible mechanism for RF processing explored in DC FE studies by Bonin at Saclay [134, 135], emission may stop abruptly if the repulsive force due to the electric field becomes stronger than the adhesive force that binds the particle to the surface, so that the emitter is ripped away from the surface and moves to a low electric field region. Emission may also *activate* abruptly during RF conditioning, presumably if a particle arrives at a high electric field region. Sensitive thermometry studies [51] show emitter turn-on associated with the arrival of a particle. Further RF processing is sometimes effective to remove the new emitters also.

### *Continuation of RF processing at higher powers*

Eventually, the effectiveness of RF processing subsides due to the limits of the available RF power, and the emission from remaining emitters becomes stable. After this, conditioning can continue if higher RF power is possible as shown in an extensive study by Graber, Crawford and Padamsee with 3 GHz and 1.3 GHz cavities [112, 133, 136]. CW power is not necessary for processing because of the short time (ns) for processing events to take place. Therefore, short pulses (μs) of high RF power can continue the processing by allowing access to even higher electric fields. In *high pulsed power processing* the power level, pulse length, and input coupling were arranged to reach progressively higher electric fields, destroying emitters along the way [132, 133]. The processing mechanism and emitter destruction effects (craters and starbursts) were shown to be the same as with CW RF processing after cavity dissection and SEM examination.

Crawford and Padamsee showed that higher power (up to one MW) short pulsed RF processing has the capability to increase the gradient of a field emission limited cavity by factors of 2 or 3 [132, 133]. The pulsed power conditioning needs to be carried out in small steps of increasing power to minimize the amount of surface damage that may accompany the destruction of active emitters. Several 5-cell strongly emitting 1.3 GHz structures *prepared without HPR* were successfully processed with RF powers from hundred kW to 1 MW and 150 μs pulse length. The field emission limited gradient was raised from 12 MV/m before processing to 26 – 28 MV/m after processing (Fig. 9.52). Note that the HPR cleaning technique had not yet been developed so that these cavities showed strong field emission before high power processing. After completion of high power RF processing the $Q$ vs. $E$ curve was limited by the high-field $Q$-slope at field levels > 25 MV/m (see next section) with some residual field emission. With 1 MW power during processing, the pulsed field accessible was as high as 90 MV/m. In all cases the low field $Q_0$ value remained at $10^{10}$, showing that the even many craters introduced during processing did not introduce significant rf losses overall, since the areas are microscopic (μm)$^2$.

The study showed that to operate field emission free at a given field level the processing field during the pulsed stage needs to reach 1.4 – 1.7 times the desired CW operating field because many emitters are active. But these studies were conducted with cavities prepared without HPR, and so the cavities had a large density of emitters present. We can expect that in the future this ratio will be smaller for cavities prepared with HPR, where the emitter density is reduced by a factor of 10 or more.

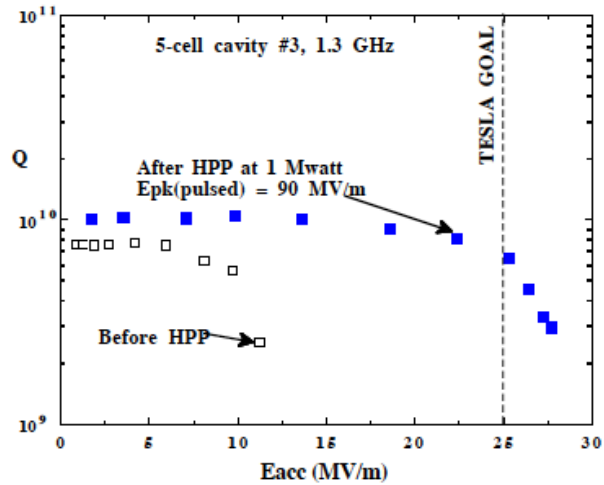

**Figure 9.52:** Effectiveness of high pulsed power processing (HPP). The 1.5 GHz, 5-cell cavity was prepared by BCP and *without* high pressure water rinsing (not yet invented). Performance was limited to 12 MV/m by heavy field emission. Field emission from many emitters was successfully processed in stages with a maximum of 1 MW of pulsed RF power after which the $Q_0$ was limited by the high field $Q$-slope (120ºC baking cure for HFQS was not yet invented). The $Q_0$ value stays at the $10^{10}$ level out to 20 MV/m when HFQS begins for the BCP treated cavity.

In the long run, HPP has the potential to be a useful technique to recover cavities which may become contaminated during assembly or during a mild vacuum accident in the accelerator. Such an experiment was conducted by Graber and Padamsee with a 9-cell 3 GHz cavity which had a best performance of $E_{acc}$ = 18 MV/m, without HPR [137]. A mild vacuum accident (few torr exposure) dropped $E_{acc}$ to 15 MV/m with heavy field emission due to particle contamination introduced during the accident. Using HPP with the available 90 kW of power it was possible to recover the original best 18 MV/m performance.

In general, RF conditioning lowers the field emission current by an order of magnitude and raises the gradient by a few MV/m depending on the amount of RF power applied. For example, at CERN, the average gradient of 350 MHz, 4-cell Nb-Cu cavities rose from 6 MV/m (design) to

7.5 MV/m by RF conditioning at 100 kW [138, 139]. Both CW and pulse conditioning with pulse length varying between 10 ms and 100 ms and duty cycle between 1 and 10% were successfully used to reduce field emission.

In retrospect, the best method for avoiding field emission is HPR rinsing after chemical surface preparation, and assembly in Class 10 – Class 100 clean rooms. However subsequent steps of cavity string assembly with introduction of couplers or other components, or vacuum accidents can re-introduce emitters, so that subsequent emitter processing methods such as He processing, plasma processing and high power RF processing become essential to recover field emission free performance.

## 9.9  High field Q-slope

After field emission was mitigated, the next barrier to high gradients arose from the "*high field Q-slope*" (HFQS). Even when there is no field emission (as judged by the absence of X-rays), the $Q_0$ starts to drop sharply above accelerating fields of 20 – 25 MV/m. (Saito [140] named the HFQS the "European Headache" around 1997.) At the time, the dominant method of chemical treatment (to etch away the damage layer of 100 – 150 microns) and final surface preparation was BCP. Its popularity was due to ease of implementation by simply dunking the cavity in an acid mixture bath. The few micron sharp steps at grain boundaries resulting from BCP did not seem to present much concern.

In 1997, Saito [140] presented a paper about the "Superiority of electropolishing over chemical polishing on high gradients". KEK had been routinely using EP for years instead of BCP for all applications, despite the complexity of the procedure involving the need for electrodes, and rotation. They believed the smoother EP surface would give better performance.

According to Fig. 9.53, Saito and his colleagues at KEK showed that if EP was used as the final chemical surface treatment for 1.3 GHz single cells, the HFQS was absent. With BCP they found gradients were always limited below 30 MV/m either by HFQS or by FE. Saito compared the performance of BCP only cavities with cavities which received BCP + EP. After EP, gradients up to 35 MV/m were reached, and in one case even 40 MV/m.

A key step carried out at KEK after both BCP and EP treatments was baking at 90ºC for 1 – 2 days, a routine procedure used mainly for thorough drying, and to obtain a good vacuum ($10^{-9}$ torr before cooldown). As realized later, Saito had serendipitously cured the HFQS with the *combination* of EP and baking. But the importance of the crucial need for the baking step came later.

In 1999, Lilje at DESY, along with colleagues at CERN and Saclay [141], showed that baking was the essential step after EP to remove the HFQS. Without baking, HFQS was still present, even in EP treated cavities, as shown in Figure 9.54. Lilje et al. also showed with the fixed thermometry discussed in Section 9.7 that the RF losses at 33 MV/m due to HFQS were predominantly in the magnetic field region of the cavity, and that baking at 120ºC substantially reduced these losses, even at the higher field of 39 MV/m. Baking at 120ºC for 48 hours allowed a single cell cavity to reach 39 MV/m.

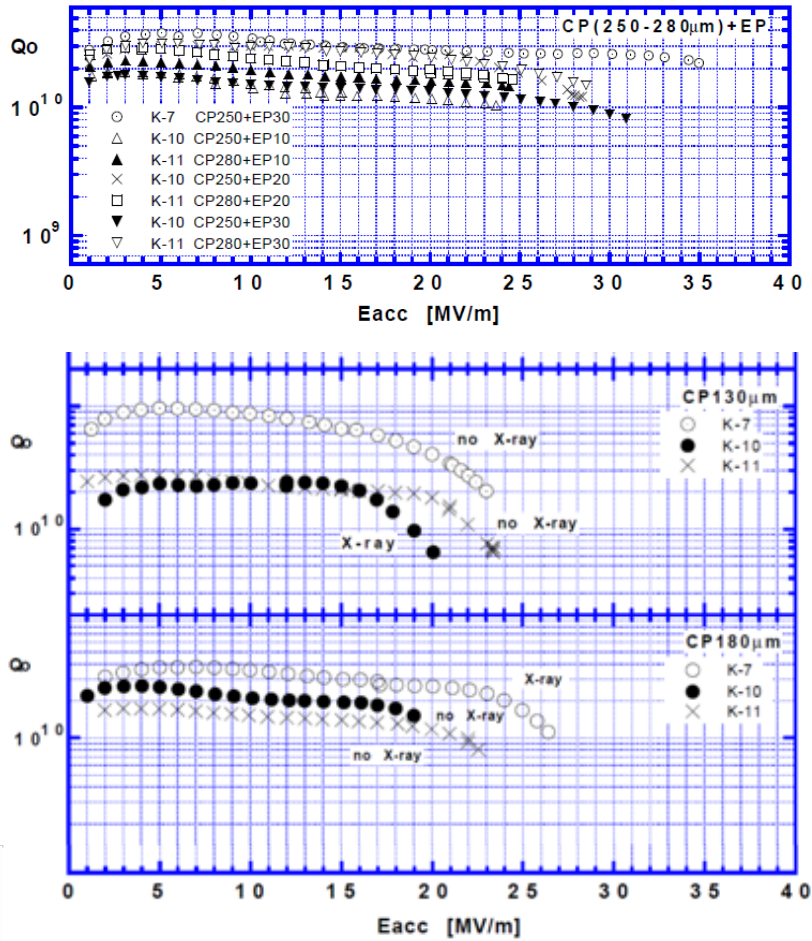

**Figure 9.53:** (Lower two curves) Single cell 1.3 GHz cavities prepared at KEK by BCP alone (130 to 180 micron removal) always showed HFQS above about 20 MV/m. (Upper) If the same cavities were treated with EP (10 – 20 microns) after BCP (250 – 280 microns) the HFQS was consistently absent.

At this stage it finally became clear why KEK was successful in taming the HFQS. Soon after, EP followed by baking became the world-wide accepted procedure for the highest gradient cavities. More than eight hundred nine-cell, 1-m-long niobium structures prepared by EP and baking have now demonstrated performance between 30 – 45 MV/m in qualification tests for the European XFEL [142], a great victory for the world-wide efforts to understand and improve SRF cavity performance.

As a pleasant surprise (Fig. 9.55), the introduction of EP and 120ºC baking also raised quench fields for cavities with lower RRR values (~200), as first discovered with single cell cavities [143] and subsequently with 9-cell cavities [2, p. 199]. A possible reason for higher quench fields with EP is the absence of sharp steps at grain boundaries abundant with BCP (Fig. 9.56). EP is well known to reduce the dimensions of surface irregularities and sharp edges are rounded as shown in Fig. 9.56 for the smoothening of the grain boundary step due to EP [144].

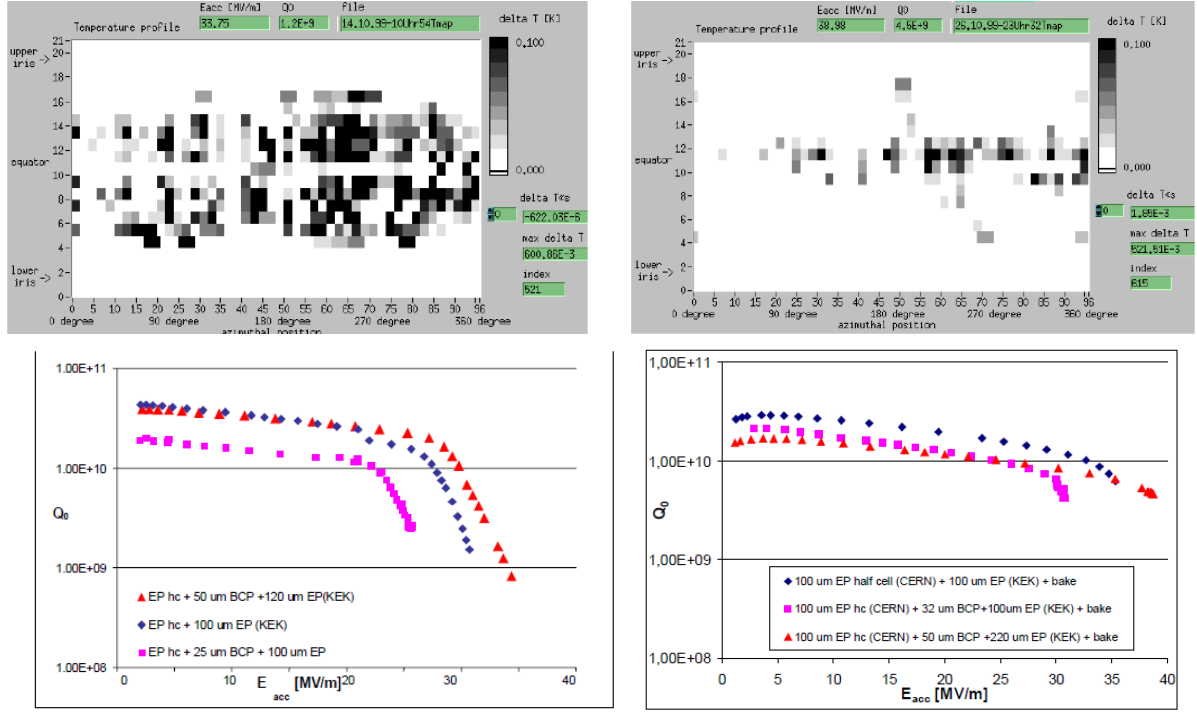

**Figure 9.54:** (Left – lower) *Q vs E* curves of 1.3 GHz single cell cavities *after EP* consistently show HFQS. (Left – upper) Temperature maps for the EP cavities with HFQS show excess heating at high magnetic field regions. (Right – lower) After baking at 120ºC for 48 hours, the HFQS disappears and gradients to 40 MV/m are possible. (Right-upper) The accompanying temperature maps of EP plus baked cavities show that the heating due to HFQS is also greatly suppressed.

When a defect is located near a grain boundary step (typically a few μm height), the local magnetic field is enhanced near the sharp step, and the cavity quench field is lower. With EP, the grain boundary steps are smaller (< 0.2 μm) and rounded – reducing field enhancement. Another reason for lower quench fields with BCP prepared surface is the higher background heating due to the HFQS.

The extra roughness of the electron beam welds due to BCP may also play a significant role in the lower quench fields. For BCP prepared cavities, the re-crystallized grains along the electron-beam weld at the equator show step heights at grain boundaries of 30 microns or more. These areas are even more likely to cause breakdown because of the greater roughness of the large grains at the weld region. By comparison, the usual grain boundary steps for BCP prepared Nb sheet is 2 – 5 microns. Moreover, the grain boundaries in the weld region are not randomly orientated, but nearly perpendicular to the magnetic field in the TM mode, which yields the largest geometric field enhancement.

Indeed, evidence for the role of electron beam welds in the quench of BCP cavities comes from cavities tested with thermometry at Saclay and KEK [143]. These BCP cavities show that quenches almost always occur at an equator seam or its vicinity. However, after electropolishing the cavity for about 50 microns or more the breakdown field increases, the *Q*-slope decreases and the breakdown location shifts to a random region in the cavity, rather than at the e-beam weld.

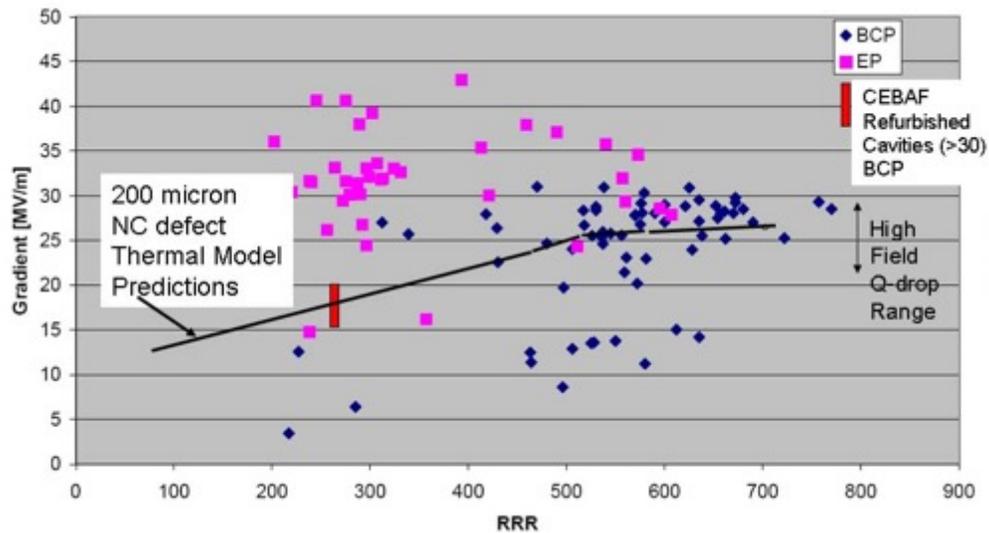

**Figure 9.55:** 9-cell cavities tested at DESY. The large scatter reflects the spread in defect size and resistance typically encountered. Data on more than 30 CEBAF re-furbished 5-cell, 1.5 GHz cavities is included [145]. Note how EP cavities show higher quench fields, even for 200 – 300 RRR Nb.

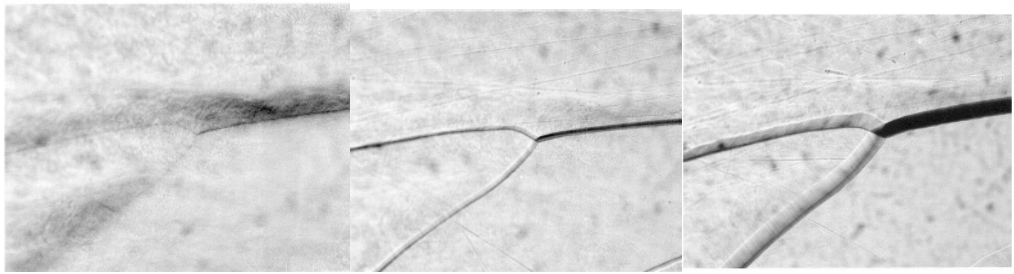

**Figure 9.56:** Progressive smoothing of a grain boundary step (from right to left) with increased amounts of EP [144].

### *The role of hydrides in the performance of niobium cavities*

While the cure for the HFQS was in hand, efforts to understand the science behind the HFQS and the baking cure procedures was still needed. In 2015, Romanenko [146] introduced the idea to explain HFQS as a milder version of the Hydrogen-related *Q*-disease. To properly discuss this topic, I need to digress to cover the history of this "disease", which affected the performance of Nb cavities in the early 1990's, as well as to cover the understanding of the disease and the cures.

In 1990, Proch at DESY [147-149] (Fig. 9.57) and Röth and Graf at Wuppertal/Darmstadt [150] discovered that when their high RRR cavities were cooled to operating temperature for the second time the *Q* values were significantly and permanently lower than when first cooled. Subsequent research over the next two years at several laboratories around the world came together to understand and cure the problem. This phenomenon is a subtle effect that depends on many factors related to hydrogen concentration in Nb. Rather than trace the chronological history of these developments (which are related mostly to $Q$), I will give a summary of the phenomena, its understanding and solution. The connection to the HFQS will become clear in the next section.

As delivered, commercial niobium typically has less than 1 wt ppm of dissolved hydrogen because the sheet material is annealed near 800°C for recrystallization. At this temperature H degasses from Nb. But hydrogen concentration in Nb cavities can increase during chemical etching, especially if the temperature of the acid etch during BCP is allowed to rise above 15°C, or if hydrogen bubbles during EP are not allowed to escape freely. H is absorbed freely by Nb when the protective oxide layer is removed by HF in the chemicals. At 10 wt ppm, hydride precipitation and high residual loss are certain for a high-purity niobium cavity, even if fast cool-down is attempted. With BCP the phosphoric acid is known to be the major culprit for H contamination. When it was replaced by lactic acid the $Q$-disease from acid etching was markedly reduced. However, this approach to solving the $Q$-disease fell out of favor as the lactic acid slowly reacts exothermically with nitric acid in BCP to cause an explosive mixture with possible consequences to personnel.

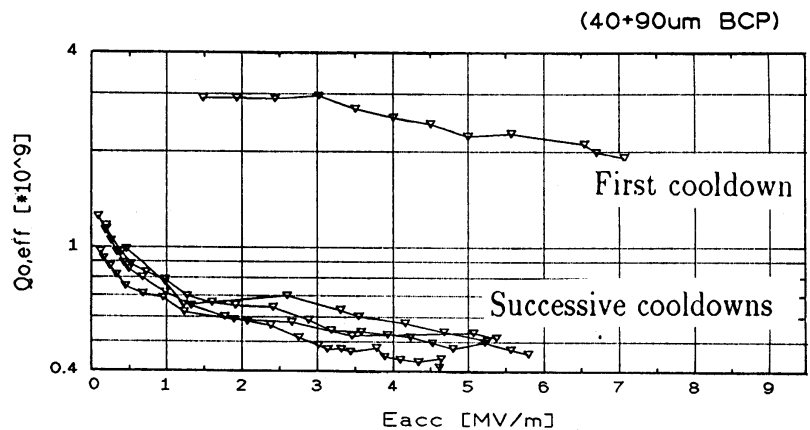

**Figure 9.57:** Irreversible behavior of the Q degradation observed at DESY with their 500 MHz cavities operated at 4.2 K.

According to the phase diagram of the Nb-H system [151], the required concentration of hydrogen to form the hydride phases is very high at room temperature (many thousands of wt ppm). Therefore, these phases do not form. As the temperature is lowered, the hydrogen concentration needed to form the hydride phases decreases. Above 150 K the danger of hydride formation is still not very serious, because the concentration required is still relatively high. Hence a cavity can be cooled as slowly as desired to 150 K. As the temperature is lowered below 150 K, the hydrogen concentration required to form the hydride phases decreases to a dangerous level, so that islands of the hydride phase may form even when the concentration is as low as 2 wt ppm. The hydride precipitates at favorable nucleation sites. If these are at the surface, they increase the residual loss. Below 150 K, the low hydride formation concentration poses a great danger for the $Q$-disease. Also, the diffusion rate of hydrogen between 150 and 60 K remains quite significant, so that hydrogen can move to accumulate to critical concentrations at hydride nucleation sites. Only when the temperature is reduced to below 60 K does the diffusion of hydrogen slow down enough that hydrogen can no longer accumulate at hydride centers.

The sharp drop in $Q$ for a cavity with strong H-disease (Fig. 9.57) indicates that Nb-H regions are initially superconducting at low fields (< 5 MV/m) and become normal at higher fields. An important aspect of the disease is that low RRR (about 30) Nb cavities do not show the *Q*-disease because interstitial impurities (such as oxygen) in Nb serve as trapping centers for hydrogen, preventing hydrogen mobility and thereby hydride growth. Similarly, vacancies and dislocations are also effective traps for hydrogen [152, 153]. Formation of Nb-H islands physically forms micro-dents on the Nb surface as H expands the Nb lattice. These dents act as nucleation sites for formation of hydrides on subsequent cool-downs, so that the *Q*-disease degradation repeats. But for the first cool down when the hydrides have not yet formed there are no dents and therefore less nucleation sites for hydride formation. Thus, the first cool-down does not show the *Q*-disease, and often gives high *Q*.

At Saclay, Antoine's measurements in 1991 of the hydrogen concentration near the surface (Fig. 9.58) [154] show that hydrogen concentrates with a large peak (as high as 5 at% = 500 wt ppm). The width of the concentration peak is about 40 nm, which is the same order of magnitude as the penetration depth in Niobium. Hence hydrogen contamination is very dangerous to the superconducting properties of Niobium. Even the sample annealed at 1000 C for several hours to degas most of the bulk H continue to show a H rich peak at the surface, probably due to H exposure from the furnace or from air. Later results [155, 156] showed much higher surface concentrations (up to 25 at. %) for chemically treated samples, as well as for annealed samples (5 – 25 at%). Also, the width of hydrogen rich region was found to be 50 nm.

The best cure found for the *Q*-disease is to degas most of the H by baking the cavity at 800°C for a couple hours in a vacuum of better than $10^{-6}$ Torr, or at baking at 600°C for 10 hours, if the decrease in yield strength from 800°C bake cannot be tolerated. Most of the bulk H is removed, but as Fig. 9.58 (b) shows, it is impossible to remove all the H, and the surface concentration effect is still present at a dangerously high level.

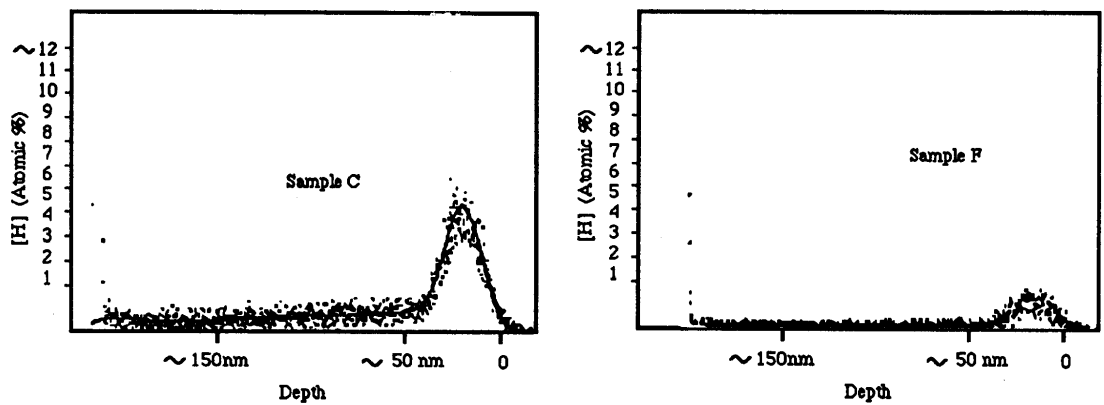

**Figure 9.58:** Elastic Recoil Detection Analysis (ERDA) for hydrogen concentration near the surface of a Nb with 200 RRR prepared by (left) BCP (right).

*HFQS model*

In addition to the hydride studies related to the hydrogen *Q*-disease, several other observations were also helpful to understanding HFQS. The baking (120°C) benefit is preserved even after the cavity is subsequently exposed to air and water. The baking benefit does not change when the oxide layer is removed (by HF rinsing) and a new layer grown. Therefore, baking improves the properties of the RF surface down to a depth of at least several nm. But the benefit does start to deteriorate progressively with repeated cycles of HF rinsing until all the benefit is gone after 5 or 6 cycles. These results showed that the baking benefit extends down to a depth of ~10 nm, as each cycle of HF rinsing removes about 1 – 1.5 nm of the surface [157].

As discussed in the previous section, a hydrogen-rich layer (estimated at several - 25 atomic %) exists near surface of SRF niobium cavities prepared by the standard methods. Even after 600 – 800ºC heat treatment removes most of the bulk H to prevent the *Q*-disease, some H re-enters the oxide-free surface from the residual H in the furnace and from subsequent exposure to clean air. H segregates in the first few tens of nanometers at the oxide–metal interface because of the local strain induced by the oxide layer and the presence of other impurities such as oxygen. Between 100 and 150 K, Nb-H precipitates form in this H-rich layer at favorable nucleation sites, such as local stress-centers, vacancies of interstitial impurities. The $T_c$ of Nb-H is near 1.5 K, but the proximity effect between Nb-H and the surrounding superconducting Nb renders the hydrides superconducting at He temperatures. The onset field of the HFQS depends on the size of the hydrides, estimated to be ≥ 10 nm. Proximity superconductivity can be sustained only up to a magnetic field value inversely proportional to the smallest dimension of the hydride ($B_c \sim 1/d$). At $B \sim 100$ mT, hydrides start to transition to the normal conducting state, starting from the largest. This is the onset of the high field *Q* slope. As the field rises, the smaller hydrides turn normal.

What about the 120ºC baking cure? For some time, it was thought that baking introduced vacancies to trap the H and prevent the formation of hydrides in the RF layer [158]. But recently, Romanenko explained [159] the 120ºC baking effect by showing the diffusion of O from the surface (Fig. 9.59(b)), which traps H, preventing H from diffusing freely to form hydrides. This is somewhat similar to the effect that low RRR Nb cavities do not show the H related *Q*-disease due to interstitals.

*Nitrogen infusion cures HFQS*

Grasselino at Fermilab in 2017 discovered another method called nitrogen infusion [160] to remove the HFQS. After 800ºC heat treatment to remove H, the temperature is lowered to 120ºC for 48 hours with 25 mtorr of N in the furnace. If no N is admitted, the HFQS is still present. But with N present during the furnace bake at 120ºC, HFQS is cured and accelerating fields above 40 MV/m are obtained in single cells at 37 MV/m in a 9-cell. Fig. 9.59(a) shows N-infusion results that are superior to the best 120ºC baking results.

Romanenko's H-trapping explanation for the 120ºC baking also applies here to how N-infusion eliminates the HFQS. His results from SIMS analysis show that the standard 120ºC bake introduces an O profile in the RF penetration depth, and the N-infusion procedure introduces a N profile. O and N impurities serve as trapping centers for H diffusion, preventing the formation of Nb hydrides with are responsible for HFQS. Apparently N interstitials provide better trapping centers for H so that higher *Q* and higher gradients are possible with N-infusion than with 120ºC baking (which can be thought of as oxygen-infusion).

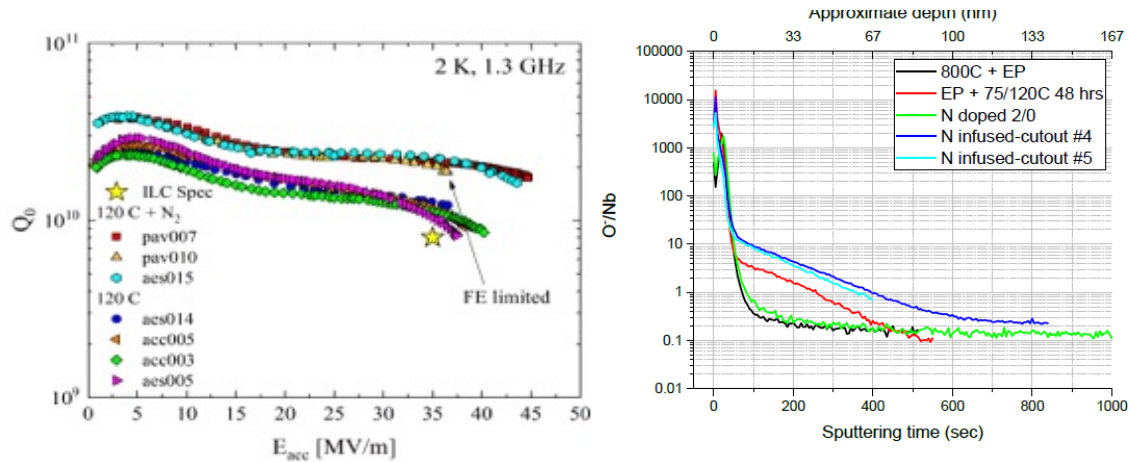

**Figure 9.59:** (a) First successes of N-infusion at Fermilab to achieve high gradients with high $Q$s. (b) SIMS analysis of 120°C baked samples show O profiles in the RF layer, and N-infusion samples show N profiles in the RF layer. These impurities serve as trapping centers for H and prevent the formation of Nb-H in the RF layer.

## 9.10 New cavity geometries

In 2001, when the record accelerating field was 42 MV/m, Shemelin and Padamsee [161] proposed a special new cavity shape (called Re-entrant) to lower $H_{pk}/E_{acc}$ by about 10 – 15% by rounding the equator to expand the surface area of the high magnetic field region [161, 162], and by allowing $E_{pk}/E_{acc}$ to rise by about 20%. The re-entrant shape has an $\Omega$-like profile with $H_{pk}/E_{acc} = 37.8$ Oe/(MV/m) and $E_{pk}/E_{acc} = 2.4$ as compared to 42 and 2.0 for the standard TESLA shape. The 20% increase in $E_{pk}$ makes cavities with the new shapes more susceptible to field emission. The motivation in trying the new shape was that quench, governed by $H_{pk}$, is a hard limit, whereas field emission, governed by $E_{pk}$, could be improved over time with better cleaning of the surface. In 2002, Sekutowicz [163] proposed the Low-Loss shape with a reduced aperture to reduce $H_{pk}/E_{acc}$ but also to reduce cryogenic losses with a higher $R/Q$ for the better shape. In 2004, Saito [164] proposed a variant of the Low-Loss shape which he named the Ichiro shape. The goal of the new cavity geometry was to demonstrate gradients over 50 MV/m. The number "50" would honor the jersey number of the famous Japanese baseball player, Ichiro. To achieve 50 MV/m gradient requires field emission free behavior to 120 MV/m, as compared to the best $E_{pk}$ achieved with standard TESLA cavities of 90 MV/m (at 45 MV/m gradient). Therefore, better surface cleaning than HPR will be necessary in the future, as well as high power RF pulse conditioning to reduce emission.

Fig. 9.60 shows several single-cell, 1.3 GHz cavities of new geometries. Single cell cavities of the new shapes demonstrated gradients of 50 – 54 MV/m with $Q_0$ values above $10^{10}$ [165, 166]. A record field of 54 MV/m at $Q$ about $10^{10}$ was set by a single cell re-entrant cavity with reduced aperture, and 59 MV/m at $Q$ about $3\times10^9$ (see Fig. 9.61) [167].

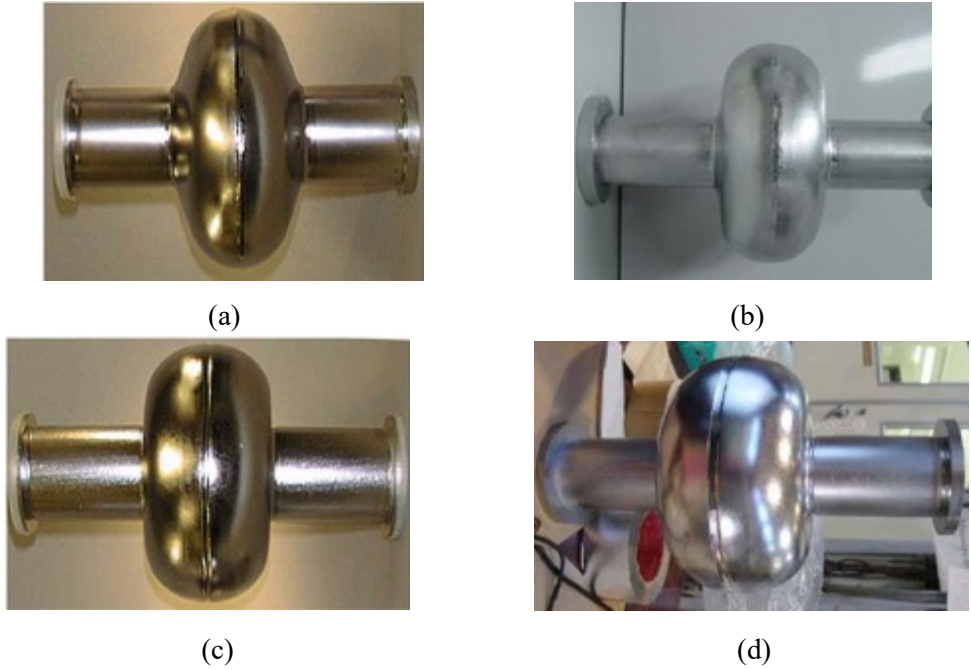

**Figure 9.60:** Single cell, 1.3 GHz niobium cavities of various shapes: (a) TESLA shape; (b) KEK, Low-Loss (ICHIRO) with 60 mm aperture; (c) Cornell Re-entrant with 70 mm aperture; (d) Cornell Re-entrant with 60 mm aperture.

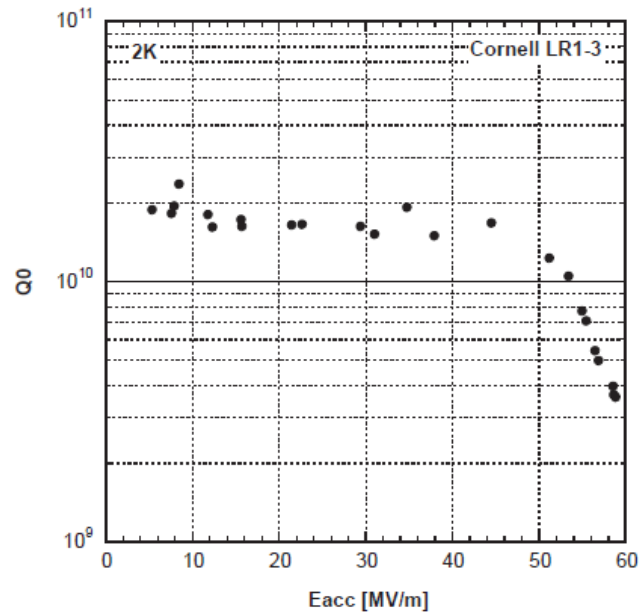

**Figure 9.61:** Record accelerating gradient achieved by the Cornell re-entrant cavity with the 60-mm aperture [167].

Multicell cavities of Re-entrant, Low-Loss and Ichiro types have been built and tested. However, the best multi-cell cavities of the new shapes have only reached 42 MV/m [4], mostly due to the dominance of field emission. The 20% higher $E_{pk}$ leads to increased field emission. The effort continues for better methods to reduce field emission by better cleaning.

A relative newcomer to the new shape effort is the LSF (Low Surface Field) shape which reduces $H_{pk}$ without raising $E_{pk}$ [168]. This shape may be the answer to the field emission limit plaguing the multicell cavities of the advanced shapes. Single cells and multicells have been fabricated and soon to be tested [169].

## 9.11 Conclusions and remarks for the future

With continued progress in basic understanding of SRF science, the performance of cavities has steadily improved to approach theoretical capabilities of niobium. In some cases, the understanding followed the invention of cures by serendipity. The major breakthroughs for gradient successes with accompanying application benefits have come from the anti-multipacting spherical cavity shape, followed by the more widely used elliptical shape, high thermal conductivity, high RRR Nb to avoid quench, high pressure (100 bar) water rinsing to quell field emission, and finally electropolishing followed by 120ºC baking to remove the high field $Q$-slope, as well as the new nitrogen-infusion method.

Chapter 4 has discussed a new two-step bake procedure [170] which reliably demonstrated a gradient of near 50 MV/m in a TESLA shape cavity with $H_{pk}/E_{acc}$ = 42 mT/MV/m. Combining nitrogen-infusion procedure or two-step bake with one of the advanced shape cavities has the potential of improving the gradients to above 65 MV/m.